\documentclass[12pt]{article}
\pdfoutput=1
\usepackage{graphicx}
\usepackage{subfigure,amssymb,amsmath}
\usepackage{cite,color,xcolor,url,cancel,ulem}
\usepackage{bm}
\usepackage{float}
\usepackage{jheppub}
\definecolor{light-gray}{gray}{0.75}
\def\321{$\rm SU(3)_C\times SU(2)_L\times U(1)_Y$}
\def\displ{{}}

\def\Vev{{\it vev}}
\def\Vevs{{\it vevs}}
\def\321{$\rm SU(3)_C\times SU(2)_L\times U(1)_Y$}

\def\10{SO(10)}
		
\def\mtl{m_{\tilde{t}_L}}
\def\mtr{m_{\tilde{t}_R}}
\def\mbr{m_{\tilde{b}_R}}

\def\stop{\tilde{t}}
\def\sb{\tilde{b}}

\def\stau{\tilde{\tau}}

\def\lspone{\widetilde\chi_1^0}
\def\mlspone{m_{\lspone}}

\def\lsim{\ ^<\llap{$_\sim$}\ }                                                
\def\gsim{\ ^>\llap{$_\sim$}\ } 
\def\amususy{a_\mu^{\rm SUSY}}
\def\gmin2{(g-2)_\mu}
\def\bsg{{\rm Br}(B \rightarrow X_s +\gamma)}
 
\def\smuon{\tilde{\mu}}
\def\issue(#1,#2,#3){{\bf #1}, #2 (#3)}
\def\PREP(#1,#2,#3){Phys.\ Rep. \issue(#1,#2,#3)}
\def\lspone{\widetilde\chi_1^0}
\def\mlspone{m_{\lspone}}

\newcommand{\tb}{\tan\beta}
\newcommand{\stopr}{\tilde{t}_R}
\newcommand{\stopl}{\tilde{t}_L}
\newcommand{\stopone}{\tilde{t}_1}
\newcommand{\stoptwo}{\tilde{t}_2}
\newcommand{\sbottoml}{\tilde{b}_L}
\newcommand{\sbottomr}{\tilde{b}_R}

\newcommand{\ms}[2]{m_{\tilde{#1}_{#2}}}

\newcommand{\MW}{M_{W}}

\title{Exploring Charge and Color Breaking vacuum in Non-Holomorphic MSSM}

\author[a, b]{Jyotiranjan Beuria}
\affiliation[a]{Harish-Chandra Research Institute, HBNI, Allahabad 211019, India.}
\affiliation[b]{Regional Centre for Accelerator-based Particle Physics \\
	Harish-Chandra Research Institute, Allahabad 211019, India.}
\author[c]{Abhishek Dey}
\affiliation[c]{Maulana Azad College, Government of West Bengal,\\ 8 Rafi
Ahmed Kidwai Road, Kolkata 700013, India.
}
\emailAdd{jyotiranjan.beuria@gmail.com}
\emailAdd{dey.abhishek111@gmail.com}
\preprint{HRI-P-17-08-003 \\ 
\vspace*{-0.8cm}
\begin{flushright}
HRI-RECAPP-2017-008
\end{flushright}
}
\abstract{Non-Holomorphic MSSM (NHSSM) shows various promising
features that are not easily obtained in MSSM. However,
the additional Non-Holomorphic (NH) trilinear interactions that attribute to
the interesting phenomenological features, also modify the 
effective scalar potential of the model significantly.  
We derive  analytic constraints involving trilinear parameters $A_t'$ and $A_b'$ 
that exclude global charge and color breaking minima (CCB).  Since
the analytic constraints are obtained considering 
specific directions in the multi-dimensional field space, 
we further probe the applicability of these constraints by exhaustive scan over NH parameter space with two different regimes of $\tan\beta$ and 
delineate the nature of metastability by considering vacuum 
expectation values for third generation squarks. We adhere to a natural scenario by fixing Higgsino mass parameter ($\mu$) to a low value and 
estimate the allowed ranges of NH trilinear parameters by considering vacuum stability and observed properties of Higgs as the determining criteria. }  
\keywords{Beyond Standard Model, Supersymmetry Phenomenology}
\begin{document}
\maketitle
\section{Introduction}
The electroweak symmetry breaking (EWSB) and strong interactions are 
quite successfully explained in the Standard Model (SM) \cite{SMrefs} 
of particle physics. The discovery 
of the SM-like Higgs boson of mass about $125$ GeV at the Large Hadron
Collider experiments at ATLAS \cite{HiggsDiscoveryJuly2012} and CMS
\cite{Chatrchyan:2013lba} marks the end of particle searches within the  SM. In such a scenario, studies Beyond the Standard Model (BSM) 
are motivated by the quest for new physics that is essential for 
addressing the puzzles that are not resolved in the SM. BSM theories are 
required to provide a potential solution to the so-called gauge
hierarchy problem along with explaining the massive neutrinos 
and providing suitable particle candidate for Dark Matter (DM)
\cite{Jungman:1995df,Bertone:2004pz,Garrett:2010hd}. Supersymmetry (SUSY)
\cite{SUSYreviews1,SUSYbook1,SUSYbook2,SUSYreviews2} is one of the most 
widely explored BSM theories.  The simplest SUSY extension of the SM, 
the Minimal Super Symmetric Standard Model (MSSM) addresses many of the 
issues that are unresolved in the SM. 

In the MSSM, the SM fermions and bosons are supplemented by bosonic and  
fermionic partners transforming under the SM gauge group  $SU(3)_C\times SU(2)_L\times U(1)_Y$. As a result,  
the particle content of the MSSM is significantly enhanced. Apart from the SM
particles, squarks ($\tilde{q}, \tilde{d}$) and sleptons ($\tilde{l}$) that 
are charged under $SU(3)_C$ and /or $U(1)_{EM}$ are present in the MSSM. Additionally, 
the fermionic counter parts of Higgs bosons and gauge bosons are 
present in the SUSY model. The color singlet higgsino and the gaugino combine 
to form charged and neutral mass eigenstates namely the charginos and the neutralinos. 
The lightest neutralino in a $R$-parity conserving scenario is a 
viable particle candidate for a cold dark matter. 
However, the absence of any hint of SUSY at the LHC has considerably constrained 
the MSSM. The requirement of large radiative corrections to
Higgs boson mass $m_{h_1}$ demands heavier stops or larger stop mixing trilinear 
soft SUSY breaking parameter ($A_t$). However, such a large $A_t$ is severely 
constrained from the $\bsg$ limits at large $\tb$ \cite{Haisch:2012re}. On 
the other hand, the well-motivated higgsino dark matter that can easily accommodate the relic abundance data from PLANCK \cite{Ade:2013zuv} and direct detection cross-section measurements from LUX \cite{Akerib:2013tjd} experiment demands the lightest SUSY particle (LSP) to be about  $1~\mathrm{TeV}$. Moreover, 
the results from Brookhaven experiment \cite{Bennett:2006fi}
 for the anomalous magnetic moment of muon or  $\gmin2$ show about $3\sigma$ deviation from the SM predictions creating a scope for a new physics explanation. In the MSSM, a very light smuon ($\smuon$) is required to accommodate the $\gmin2$ results within  $1\sigma$, in a region where the SUSY contribution $(\amususy)$ \cite{Kowalska:2015zja,Heinemeyer:2003dq,
		Chattopadhyay:2000ws,Moroi:1995yh,Chattopadhyay:1995ae,Lopez:1993vi,
		Jegerlehner:2009ry} is dominated by the $\smuon-\lspone$ loops. On the contrary, 
$\smuon$ with very low mass is not so friendly with the LHC data \cite{ATLAS:2017uun,Sirunyan:2017lae}.

Apart from the  experimental and phenomenological aspects, there are a few theoretical issues that need to be explored. 
In view of current experimental constraints, ElectroWeak Fine Tuning (EWFT) is a 
major concern in the phenomenological MSSM (pMSSM). 
In general, larger value of the bilinear Higgs mixing parameter of the superpotential  $(\mu)$ significantly enhances the EWFT \cite{Mustafayev:2014lqa}. 
In the MSSM, even in scenarios that are devoid of any Higgsino like LSP, 
lower bound on $|\mu|$ is around $100$ GeV via LEP limits on lighter chargino.  
Moreover, in a Higgsino DM setup,  relic density limits from PLANCK constrain $\mlspone$ to be $\gsim$ $1$ TeV, that in turn necessitate $|\mu|\sim 1$ TeV  enhancing the 
EWFT significantly.

It has been 
shown in Refs \cite{Un:2014afa2,Chattopadhyay:2016ivr} that extension of the 
generic MSSM via the inclusion of non-holomorphic (NH) soft SUSY breaking 
terms, quite satisfactorily ameliorates all the above problems of the MSSM. 
This extended MSSM, namely the Non-Holomorphic MSSM (NHSSM) has the particle 
content identical to that of the MSSM, but have extra soft SUSY breaking
terms. The NH higgsino mass parameter $\mu'$ present in the NHSSM,
contributes to the Higgsino content of the neutralinos, but does not
have any signature on the tree level neutral scalar potential. The
NHSSM can accommodate Higgsino DM even for very low value of $\mu$
resulting in the reduction of EWFT \cite{Mustafayev:2014lqa,Ross:2016pml, Chattopadhyay:2016ivr,Ross:2017kjc}. 
The tri-linear NH parameters control the L-R mixing of corresponding squarks and sleptons.
Apart from the masses, the L-R mixing effect affects various processes
that may be relevant phenomenologically. In this analysis we will focus
on the trilinear NH terms for their roles in influencing the scalar
potential of the theory, particularly for the possible appearance of
 charge and color breaking (CCB) minima    
 \cite{Casas:1995pd,LeMouel:2001ym,LeMouel:2001sf,AlvarezGaume:1983gj,
Gunion:1987qv,Strumia:1996pr,Baer:1996jn,Abel:1998cc,Abel:1998wr,Ferreira:2004yg,
Brhlik:2001ni,Ferreira:2000hg,Bordner:1995fh,Cerdeno:2003yt}.  Some well-known analytic 
constraints with simplified assumptions have been used in the MSSM in order to avoid the regions of parameter 
space where CCB minima is deeper than the desired symmetry breaking (DSB)
SM like (SML) vacuum. 
Consideration of scenarios with global CCB minima, where the DSB minima are stable with respect to tunneling to 
the deeper CCB minima have led to considerable relaxation of the analytic CCB  constraints in 
the MSSM \cite{Chattopadhyay:2014gfa,Chowdhury:2013dka}.
If local DSB minimum has a large lifetime in regard to quantum tunneling to  
global CCB minima, the corresponding DSB minima is referred to as `long-lived' SML vacuum in the 
literature \cite{Riotto:1995am,Falk:1996zt,Kusenko:1996jn,Kusenko:1996xt,Kusenko:1996vp,Kusenko:1995jv, Brandenberger:1984cz}. 
The long-lived and absolutely stable DSB minima are often collectively referred to as safe vacuum.  
While exploring the phenomenological features of a model one should consider only those regions 
of parameter space that are associated with safe vacuum. Besides the above,
inclusion of finite temperature effects \cite{Brandenberger:1984cz}  significantly modifies the vacuum 
structure further and thus, constraining the allowed region of parameter space with respect to the 
stability against thermal tunneling. Analyses of vacuum stability considering both 
thermal and quantum mechanical instabilities have been done in popular models the  
MSSM \cite{Camargo-Molina:2013sta,Camargo-Molina:2014pwa, Bobrowski:2014dla, Hollik:2015pra, Hollik:2016dcm, Beuria:2017wox} 
and the NMSSM \cite{Ellwanger:1999bv, Beuria:2016cdk,Krauss:2017nlh,Beuria:2017wox}.  

In this work, we would probe the vacuum structure of the NHSSM and study the stability of DSB minima 
using our implementation of complete 1-loop corrected scalar potential in {\tt Mathematica} and the publicly 
available package $\tt{Vevacious}$ \cite{Camargo-Molina:2013qva} which in turn uses {\tt CosmoTransitions} \cite{Wainwright:2011kj} 
for determining tunneling time to a deeper vacuum. We would determine the analytic constraints to avoid global CCB vacua using 
tree-level scalar potential and explore the extent of its applicability. We shall also study the role of different NH parameters 
in radiative correction to $m_{h_1}$ in order to identify phenomenologically important regions of NH trilinear SUSY breaking parameters 
associated with third generation squarks sector {\it viz.} $A_t'$ and $A_b'$ respectively.

The analysis is structured as below.
In Sec.\ref{NHSSM} we very briefly introduce the NHSSM and discuss the analytical CCB constraints 
and dependence of effective potential on tri-linear NH parameters. We also analyze the role of 
vacuum expectation values (\Vevs) of stops ($\stop$) and sbottoms ($\sb$) in determining the fate of the DSB vacuum while 
satisfying the Higgs mass constraints. In Sec.\ref{results} we present the results and analyze 
them in the light of theoretical predictions. Finally, we conclude in Sec.\ref{conclusion}.
\section{Non-Holomorphic MSSM}
\label{NHSSM}
Going beyond the generic MSSM, NH SUSY breaking  terms potentially fall 
in the class of terms that may cause hard SUSY breaking \cite{grisaru,
	Martin:1999hc,Haber:2007dj,Bagger:1993ji,Ellwanger:1983mg,Jack:1999ud,Frere:1999uv,Martin:2015eca}, 
particularly in models with singlet scalars. The ``may be soft''\cite{Martin:1999hc}  D-term contributions like  $ {1\over M^3}[ X X^* \Phi^2 \Phi^* ]_D$ and  
$ {1\over M^3}[ X X^* D^\alpha \Phi D_\alpha \Phi ]_D$ give rise to the NH terms in the Lagrangian {\it viz.} $\phi^2 \phi^* $ and $\psi\psi $ respectively. Here $X$ and $\Phi$ are
chiral superfields and SUSY breaking in the hidden sector is 
driven by $\Vev$ of an auxiliary field $F$ belonging to $X$.
We note that $<F>/M$ should refer to a weak scale mass which we consider here 
as the W-boson mass $M_W$.  However, in models like pMSSM it  can be varied 
independently. Hence, coefficients of $\phi^2 \phi^* $ and $\psi\psi $ varies as $\displ {|<F>|^2 \over M^3} \,\sim\,{M_W^2 \over M}$. Thus, it is evident that they are highly suppressed in super-gravity  type of models. However, 
in weak scale scenario $\phi^2 \phi^* $ and $\psi\psi $ are quite significant and appear as the NH trilinear terms and bare Higgsino mass term respectively
\footnote{For details discussion see Ref.\cite{Chattopadhyay:2016ivr} and references therein.}.

This was discussed or at least pointed out in several works \cite{Martin:1999hc,
Haber:2007dj,Bagger:1993ji,Jack:1999ud,Jack:1999fa,Hetherington:2001bk,
Jack:2004dv,Cakir:2005hd,Un:2014afa1}.  Impact of NH parameters on $m_{h_1}$,   
electroweak fine tuning, $\bsg$ and $\gmin2$ have been analyzed in Ref.\cite{Chattopadhyay:2016ivr}. 
Earlier works involving the analysis of the effects of NH terms include Ref.\cite{Solmaz2009,Hambye:2000zs,SRoy}. Here we would like to explore the scalar 
potential of NHSSM with emphasis on the vacuum stability issues.  Our prime 
focus will be on the attribution of the NH terms towards determining the 
fate of the EWSB  vacuum.
\subsection{CCB in NHSSM}
\label{ccbnhssm}
We remind that the MSSM is considered to have only holomorphic soft 
SUSY breaking terms. The trilinear soft terms, in particular,  
are given by as follows \cite{SUSYbook1}.
\begin{equation}
\label{h_lagrangian}
-\mathcal{L}_{soft}\supset \tilde{Q}\cdot H_u y_t A_t\tilde{U} + 
\tilde{Q}\cdot H_d y_b A_b\tilde{D} +\tilde{L}\cdot H_d y_{\tau} 
A_{\tau}\tilde{E} +h.c. 
\end{equation}
We have only shown here the dominant terms involving the third 
generation of sfermions.  It has been shown that in the absence 
of any gauge singlet it is possible to extend the SUSY breaking 
soft sector by including NH soft SUSY breaking terms, without 
aggravating any quadratic divergence \cite{grisaru,Bagger:1993ji,
Hetherington:2001bk}.  Thus, the NH soft terms of the NHSSM in 
general that include trilinear coupling terms as well as a NH 
higgsino mass $(\mu')$ term are given by \cite{Un:2014afa1,
Un:2014afa2}
\begin{equation}
\label{nh_lagrangian}
-\mathcal{L'}_{soft}\supset \tilde{Q}\cdot H_{d}^c y_t A_{t}'\tilde{U} + 
\tilde{Q}\cdot H_{u}^c y_b A_{b}'\tilde{D} +\tilde{L}\cdot H_{u}^c 
y_{\tau} A_{\tau}'\tilde{E} +  \mu '\tilde{H_u}\cdot \tilde{H_d} +h.c.
\end{equation}
Thus, the scalar potential at the tree level including the Higgs 
and the stop fields read as follows.
\begin{eqnarray}
\label{eq:scalrpot1}
\mathrm{V}|_{\mathrm{tree}} & = & m_2^2H_u^2 + m_1^2H_d^2 + 
m_{\stopl}^2\stopl^2 + m_{\stopr}^2\stopr^2 
-  2B_{\mu}H_dH_u + 2y_t A_tH_u\stopr\stopl \notag  \\ 
 &  & - 2y_t(\mu+A_t')\stopl\stopr H_d 
  + y_t^2(H_u^2\stopl^2 + H_u^2\stopr^2 +
 \stopr^2\stopl^2) + \frac{g_1^2}{8}(H_u^2 -H_d^2 +  \frac{\stopl^2}{3} -
 \frac{4\stopr^2}{3})^2 \notag \\  
 & & + \frac{g_2^2}{8} (H_u^2 -H_d^2 - \stopl^2)^2
 +\frac{g_3^2}{6}(\stopl^2-\stopr^2)^2,
\end{eqnarray}
where $m_2^2=m_{H_u}^2+\mu^2$ and $m_1^2=m_{H_d}^2+\mu^2$ and 
$m_{H_u}^2$ ($m_{H_d}^2$) is the soft squared mass term for 
up (down) type Higgs.
Relevant terms may be added to the above equation to construct the complete 
scalar potential at the tree level taking into account the other sfermions. 
In the above expression, vanishing value of $A_t'$ leads to the generic MSSM
scenario. In this analysis we will explore the CCB constraints at tree-level 
for deriving simple analytic result. The results from our numerical scans are obtained
via {\tt Vevacious} \cite{Camargo-Molina:2013qva} which includes 1-loop
corrected potential both at zero temperature and finite temperature. 
Furthermore, in presence of CCB minima, it computes the tunneling rate from DSB vacuum to 
the former using {\tt CosmoTransitions} \cite{Wainwright:2011kj}. However,
we consider the scalar potential only at the tree-level 
in order to obtain the analytic CCB constraints analogous to those used in the MSSM.
One loop correction (in the $\overline{DR}$-scheme) \cite{Quiros:1999jp, Martin:2014bca})
at zero temperature is given by
\begin{equation}
\Delta V_\mathrm{rad.corr.}  
=  \sum_{i} \frac{n_i}{64 \pi^2} m_i^4 (\ln \frac{m_i^2}{Q^2}-\frac{3}{2}) \;\; ,
\label{eq:radcorr}
\end{equation}
where the sum runs over all real scalars, vectors and Weyl fermions 
that are present in the model with
\begin{equation}
n_i  =  (-1)^{2s_i} (2s_i+1) Q_i C_i 
\end{equation}
and $Q_i=2 (1)$ for charged particles (neutral particles), $C_i$ is the 
color degrees of freedom, $s_i$ is the spin of the particle, $m_i$ is the mass 
of the same and $Q$ is the renormalization scale used.

The stability of the DSB vacuum can be significantly 
affected by thermal corrections to the
scalar potential. The 1-loop thermal correction to the potential at temperature `T' is 
given by \cite{Brignole:1993wv}
\begin{equation}
\label{eq:pot-thermal}
\Delta V_\mathrm{thermal} = \frac{1}{2\pi^2}\sum{ T^4 J_{\pm}\left(m^2/T^2\right)} ,
\end{equation}
where
\begin{equation}
J_{\pm}(r) = \pm \int_{0}^{\infty} dx ~x^2 \ln(1 \mp e^{-\sqrt{x^2+r}}).
\label{eq:jpm}
\end{equation}
The sum in equation \ref{eq:pot-thermal} runs over all degrees of 
freedom that couple to the scalar fields including the scalar fields themselves.  
$J_{+}$ ($J_{-}$) corresponds to the corrections arising from
bosons (Weyl fermions) and $m$ is the mass of the corresponding particle 
at zero temperature. $J_{\pm}(\frac{m^2}{T^2})$ asymptotically approaches zero 
$(-\infty)$ as $\frac{m^2}{T^2}$ approaches $~\infty$ (zero). 
It is to noted that thermal corrections would always lower the potential \cite{Camargo-Molina:2013qva,
Camargo-Molina:2014pwa} depending on the 
magnitude of $\frac{m^2}{T^2}$.
The complete scalar potential at one loop that includes both zero temperature and thermal
corrections, is given by
\begin{equation}
\mathrm{V}= \mathrm{V}|_{_\mathrm{tree}} 
+ \Delta V_\mathrm{rad.corr.} 
+ \Delta V_\mathrm{thermal} \quad .
\label{eq:full-pot}
\end{equation}
 
Analytic CCB constraints in MSSM were derived along particular direction 
of field space, namely ``D-flat'' direction \cite{Casas:1995pd}. 
Keeping this in mind, we consider non vanishing \Vevs~ for the two higgs 
scalar and  the stops fields. The latter are responsible for the 
generation of CCB minima. In the direction where
$|H_d|=|H_u|=|\stopr|=|\stopl|=\zeta$, the scalar potential (Eq.\ref{eq:scalrpot1}) reduces to\\
\begin{equation}
\label{dr1}
\mathbf{V}_\mathrm{tree} = a\zeta^4 + b\zeta^3 + c\zeta^2, 
\end{equation} 
where $a=\left(3y_t^2-\frac{g_1^2+g_2^2}{8}\right)$, 
$b=[2y_tA_t - 2y_t(\mu+A_t')]$ and 
$c=(m_1^2 + m_2^2 + m_{\stopr}^2 + m_{\stopl}^2 - 2B_{\mu})$. \\ 
Minimizing the above potential with respect to $\zeta$  and 
considering non-vanishing $\zeta$, we obtain
 $4 a\zeta^2 + 3b \zeta + 2 c = 0$.\\
Hence, the value of the field at the minima is 
\begin{equation}
\zeta= \frac{-3b\pm\sqrt{9b^2-32ac}}{8a}.
\end{equation}

From the reality condition of the roots, we get $9b^2 > 32ac$.  $V_{min} > 0$ at 
the minima implies, $a\zeta^2 + b\zeta + c > 0$.  Here we consider
a scenario with 4 \Vevs~ otherwise, the contribution of $A_t'$ and $A_t$  
cannot be taken into account simultaneously.  One may also consider the scalar 
potential excluding the holomorphic trilinear terms in a three \Vevs~ scenario 
with non vanishing \Vevs~ for $H_d$, $\stopr$ and $\stopl$ with $H_u=0$. 
Nevertheless, $A_t$ plays a very crucial role in providing adequate $m_{h_1}$.  
Besides, both the Higgs fields have non-vanishing \Vevs~ in phenomenologically
viable region of MSSM.  As a result for any realistic scenario in MSSM and beyond one has
$H_d\neq 0$ and $H_u\neq 0$.  This necessitates the consideration of non-zero \Vevs~ for $H_d$ and $H_u$.  
Using the reality condition for the $\zeta$ at the minima, in simplified scenario
with $H_d=0$ \cite{Chowdhury:2013dka} we obtain
\begin{equation}
A_t^2 > 2.67 (m_2^2 + m_{\stopl}^2 + m_{\stopr}^2),
\end{equation}
where $m_2^2 = m_{H_u}^2 + \mu^2$.  This is referred as the condition for the 
existence of a deeper CCB in MSSM  which is often approximated as
\begin{equation}
\label{ccbstmssm}
A_t^2 > 3 (m_2^2 + m_{\stopl}^2 + m_{\stopr}^2),
\end{equation}
that is the traditionally used CCB constraint associated with stop $(\stop)$ scalars receiving $\Vevs$
\cite{Casas:1995pd, AlvarezGaume:1983gj,Gunion:1987qv,Strumia:1996pr,Baer:1996jn,
Brhlik:2001ni,LeMouel:2001ym,Bordner:1995fh}.  Similar constraints are used for 
different sfermion fields. In NHSSM, we consider the following 
inequality resulting from reality condition $9b^2 > 32ac$ that implies
\begin{equation}
\label{intermediate1}
9\times 4 y_t^2\left [A_t-(\mu + A_t')\right ]^2 > 32 
\left\{3y_t^2-\frac{g_1^2+g_2^2}{8}\right\}\left (m_1^2 + m_2^2 +  
m_{\stopl}^2 + m_{\stopr}^2  -2B_{\mu}\right).  
\end{equation}
In contrast to MSSM, here we do analyze the scalar potential 
considering non-vanishing \Vevs~ for all the four scalar fields. 
The Eq.\ref{intermediate1} simplifies to
\begin{equation}
\label{intermediate2}
\left[A_t-\left(\mu +A_t'\right) \right]^2 > 
3 \left\{1-\frac{g_1^2+g_2^2}{24y_t^2}\right\}
\left( m_1^2 + m_2^2 + 
m_{\stopl}^2 + m_{\stopr}^2  -2B_{\mu}\right).
\end{equation}
Eq.\ref{intermediate2} identifies regions of parameter space associated with
global CCB minima. Neglecting the contributions from $g_1^2$ and $g_2^2$ 
with respect to $y_t^2$, the CCB constraint in NHSSM for $\stop$ fields becomes
\begin{equation}
\label{nhssmccb1}
\left[A_t-\left(\mu +A_t'\right) \right]^2 < 
3 \left( m_1^2 + m_2^2 + 
m_{\stopl}^2 + m_{\stopr}^2  -2B_{\mu}\right).
\end{equation}
One can always retrieve the traditional constraint in MSSM (Eq.\ref{ccbstmssm}) 
from Eq.\ref{nhssmccb1} assuming $A_t'=0$ and $H_d=0$.  As discussed in Ref.\cite{Casas:1995pd}, the most stringent 
CCB constraint is obtained when the relative sign between the trilinear terms in the potential
are always considered positive. Thus, in NHSSM the analytic condition predicting an 
absolutely stable DSB vacuum is 
\begin{equation}
\label{ccbconst}
\left\{|A_t|+|\mu| + |A_t'| 
\right\}^2 < 3 \left( m_1^2 + m_2^2 + 
m_{\stopl}^2 + m_{\stopr}^2  -2B_{\mu}\right).
\end{equation} 

Along with the absolute stable DSB minima, we also consider 
long-lived DSB minima, in presence of global CCB vacuum
 that in  turn will increase the valid parameter space.
Here one should note 
that rate of tunneling from DSB false vacuum to such CCB true vacuum 
is roughly proportional to $e^{-a/y^2}$, where $a$ is 
a constant of suitable dimension that can be determined via 
field theoretic calculations and $y$ is the Yukawa coupling. 
The tunneling rate is enhanced for large Yukawa couplings
\cite{Kusenko:1996jn,Kusenko:1996xt,Kusenko:1996vp,Kusenko:1995jv,
Brandenberger:1984cz,LeMouel:2001ym}. As a result, the third 
generation of sfermions will be the most important candidate in 
connection with the presence of potentially dangerous global
minima. In models like the MSSM and the NHSSM, the Yukawa couplings
vary with $\tb$ which is given
by $\tb=\frac{v_u}{v_d}$. However, unless $\tb$ is very large  
the relation $y_t > y_b > y_{\tau}$ is by and large satisfied,
consistent with the mass hierarchy of the corresponding fermions.
Hence, global CCB minima associated with $\stop$ fields will be most dangerous
due to comparatively larger $y_t$.  The $\sb$ fields may also play an important
role in determining the fate of the DSB minima. A large value of the NH
parameter $A_b'$ associated with $\sb$ significantly modifies $m_{h_1}$ 
and $\sb$ phenomenology particularly for large $\tb$. In the phenomenologically
interesting regions of parameter space,  the structure of the 
scalar potential is modified by the presence of large $A_b'$ and in such a setup, vacuum
stability needs to be re-explored. In this context, one should
study the analytic CCB constraint associated with the $\sb$
fields. Following a similar method as adopted for 
$\stop$, one gets the CCB constraint for $\sb$ to be 
\begin{equation}
\label{ccbconsb}
\left\{|A_b|+|\mu| + |A_b'| 
\right\}^2 < 3 \left\{1-\frac{g_1^2+g_2^2}{24y_b^2}\right\}\left( m_1^2 + m_2^2 
+ m_{\sbottoml}^2 + m_{\sbottomr}^2  -2B_{\mu}\right).
\end{equation}
Unlike the case of $\stop$, here we cannot neglect $g_1$ and $g_2$ with respect 
to $y_b$, since $y_b$ is not very large. In the limit of MSSM, in three  \Vevs~ scenario 
with non-vanishing \Vevs~ for $H_d$, $\sbottoml$ and $\sbottomr$, with $H_u=0$, 
we get the well-known CCB constraint for $\sb$, namely 
\begin{equation}
\label{ccbmssmsb}
A_b^2 < 3 (m_1^2 + m_{\sbottoml}^2 + m_{\sbottomr}^2),  
\end{equation}
directly from the Eq.\ref{ccbconsb}. As pointed out earlier the terms associated with $g_1$ and $g_2$ contribute 
to the potential in the direction where all the \Vevs~ are equal only if all the four \Vevs~ are non-vanishing. Thus, they 
naturally do not appear in the traditional CCB constraints, that are essentially 
derived in a simple three \Vev~ setup.  Similar analysis may be done for $\stau$ 
fields. However, we focus only on the effect of considering $\stop$ and $\sb$ \Vevs, since $y_{\tau}$ is small. At this point, before going for a detailed numerical study we try to estimate the limitations of the 
analytically derived CCB constraints of NHSSM namely Eqs.\ref{ccbconst} 
and \ref{ccbconsb}. First, the contributions of 
the $\stop$ and $\sb$ fields are considered separately.  On the other hand, in a realistic scenario
they jointly affect the vacuum structure of the model. Secondly, 
only the direction where equal \Vevs~ are attributed to all the fields is taken into account. Thirdly, in a phenomenological 
analysis of the pMSSM, $\tb$ varies over a wide range. Thus, $y_t$ and $y_b$
vary and $g_1$ and $g_2$ may not be always negligible with respect to Yukawa couplings.
Lastly, this work goes beyond an absolute vacuum stability  by accommodating long-lived scenarios, demanding a detail numerical analysis. In our work using {\tt Vevacious} \cite{Camargo-Molina:2013qva} we probe
the vacuum structure to find the DSB minima.
We explore a four \Vev~ scenario considering non-vanishing 
\Vevs~ for $\stopl$,$\stopr$ along with $H_d$ and $H_u$. In the other part of our analysis 
we consider a different four \Vev~ combination involving $\sbottoml$, $\sbottomr$, 
$H_d$ and $H_u$, in order to gauge the role of $\sb$ fields in determining the 
fate of the DSB minima.  In both the above cases, we try to estimate the  applicability of the analytic constraints. Finally, we analyze a more 
involved scenario with six \Vevs~ where  non-zero \Vevs~ were attributed to 
$\sbottoml$ and $\sbottomr$ along with  $\stopl$, $\stopr$, $H_d$ and $H_u$.
Additionally, a varying $\tb$, changes the values of $v_u$ and $v_d$ that
in turn modify the yukawa couplings. The vacuum structure thus depends on
$\tb$. We consider two different zones of $\tb$ to explore the dependency. 
\subsection{NH parameters and their impact on $m_{h_1}$}
\label{nhmhbsg}	   
Besides affecting vacuum stability, the NH trilinear parameters play a significant  
role in phenomenology where left-right mixing of sfermions are important. 
Additionally, the NH higgsino mass parameter $\mu'$ is important in processes 
dominated by higgsino and it controls the overall higgsino phenomenology. The NH parameters significantly modify the $\bsg$ and 
radiative corrections to $m_{h_1}$. The role of $A_t'$ together with $\mu'$ in 
the radiative corrections of $m_{h_1}$ has been studied in Ref.\cite{Chattopadhyay:2016ivr} 
in the low scale NHSSM (pNHSSM). Here we would like to probe 
the contribution of the NH parameters towards the radiative corrections to $m_{h_1}$ 
in order to identify phenomenologically interesting regions of NH parameters for 
exploring vacuum stability. The radiative correction to $m_{h_1}$ due to 
stop-loop is given by
\begin{align}
\Delta m_{h,top}^2= \frac{3 g_2^2 {\bar m}_t^4}{8 \pi^2 \MW^2} 
\left[\ln\left(\frac{\ms{t}{1} \ms{t}{2}}{{\bar m}_t^2}\right) + 
\frac{X_t^2}{\ms{t}{1}\ms{t}{2}} 
\left(1 - \frac{X_t^2}{12\ms{t}{1}\ms{t}{2}} \right) \right]. 
\label{stop_loop} 
\end{align}
Here $X_t=A_t-(\mu+A_t')\cot\beta$ and ${\bar m}_t$ stands for the 
running top-quark mass that includes electroweak, QCD and SUSY QCD 
corrections \cite{Pierce:1996zz}.  Apart from this, particularly for large 
$\tb$, $A_b'$ the NH trilinear soft parameter associated with the bottom squarks, 
has significant bearing on $m_{h_1}$. The contribution from 
sbottom-loop is 
\begin{align}
\label{sbottomeqn}
\Delta m_{h,bottom}^2= \frac{3 g_2^2 {\bar m}_b^4}{8 \pi^2 \MW^2} 
\left[\ln\left(\frac{\ms{b}{1} \ms{b}{2}}{{\bar m}_b^2}\right) + 
\frac{X_b^2}{\ms{b}{1}\ms{b}{2}} 
\left(1 - \frac{X_b^2}{12\ms{b}{1}\ms{b}{2}} \right) \right],
\end{align}
where $X_b=A_b-(\mu + A_b')\tan\beta$\footnote{A similar result for the stau
contribution would involve $X_\tau=A_\tau-(\mu +A_{\tau}')\tan\beta$.}.  
In the MSSM, the contribution from $\sb$  loops is important only for 
large $\tb$ and $\mu$, since ${\bar m}_b$ is significantly smaller 
than ${\bar m}_t$. Hence, in a MSSM scenario motivated by low EWFT, 
where $\mu$ is predominantly very small, the contribution to $m_{h_1}$ 
due to the $\sb$ is negligible, although the effect is $\tb$ enhanced.  
The situation is quite different in the NHSSM. Here for a large value 
of $A_b'$, $\Delta m_{h,bottom}^2$ is albeit small but not negligible 
compared to $\Delta m_{h,top}^2 $ because of the associated $\tb$ enhancement.  Thus even in a 
scenario motivated by low EWFT, the radiative correction to $m_{h_1}$ from $\sb$ 
loops can be non-negligible in the NHSSM.  However, the contribution may be 
negative or positive depending on the value of $X_b$.  
\section{Results}
\label{results}
We explore the nature of the scalar potential in a semi-analytic approach that demonstrates the dependency on 
NH parameters. We study the degree of applicability of the 
analytic CCB constraints {\it viz.} Eqs. \ref{ccbconst} and \ref{ccbconsb} in the NHSSM. Considering the 
contributions of $A_t'$ and $A_b'$ to $m_{h_1}$, we analyze 
the stability of the DSB minima using {\tt Vevacious} \cite{Camargo-Molina:2013qva}. 
 {\tt SARAH} \cite{Staub:2012pb,Staub:2013tta,Staub:2015iza,Staub:2016sms} and 
 {\tt SPheno} \cite{Porod:2003um,Staub:2009bi,Staub:2010jh,Porod:2011nf,Porod:2014xia} were used for 
model building and spectrum generation respectively. The spectra 
generated by {\tt SPheno} include complete two-loop corrections to
Higgs mass even from NH parameters \cite{Goodsell:2014bna, Goodsell:2015ira}.
The numerical study  is performed in three different set-ups of
non-vanishing \Vevs. First, we exclusively probe the stability 
of DSB minima in presence of global CCB minima associated with 
non-zero \Vevs~ of $\stopl$ and $\stopr$. Here, we consider 
a four-\Vev~ scenario with non-zero \Vevs~ for
$H_d$, $H_u$, $\stopl$ and $\stopr$. Then, we explore the
role of $\sbottoml$ and $\sbottomr$ in rendering the DSB
minima unstable. Finally, we analyze a combined multiple-\Vev~
scenario with non-vanishing \Vevs~ for six scalar fields viz.
$H_d$, $H_u$, $\stopl$,  $\stopr$, $\sbottoml$ and
$\sbottomr$.  $y_t$ and $y_b$ will significantly 
affect the vacuum stability against CCB minima. Hence, we divide 
our analysis into two $\tb$ regimes, viz.
$5 \leq \tb \leq 10$ and $40 \leq \tb \leq 50 $.  
\subsection{Nature of the NHSSM scalar potential}
\begin{figure}[t]
	\begin{center}
		\subfigure[]{%
			\label{fig:pot1a}
			\includegraphics[width=0.51\textwidth]{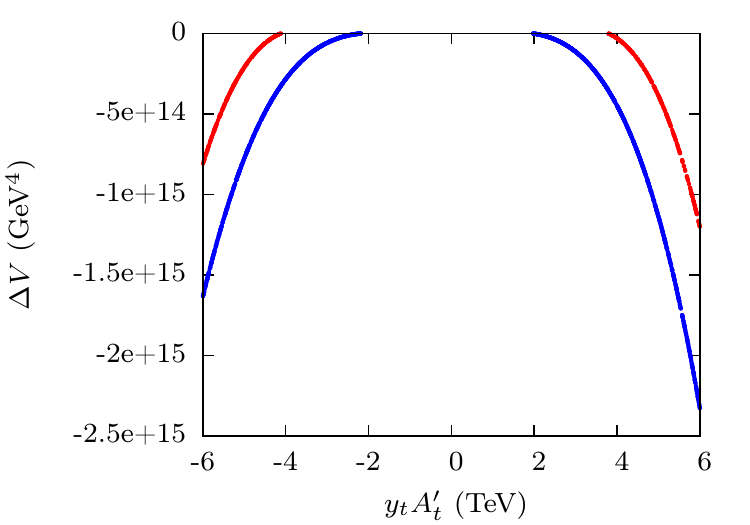}
		}%
		\subfigure[]{%
			
			\label{fig:pot1b}
			\includegraphics[width=0.51\textwidth]{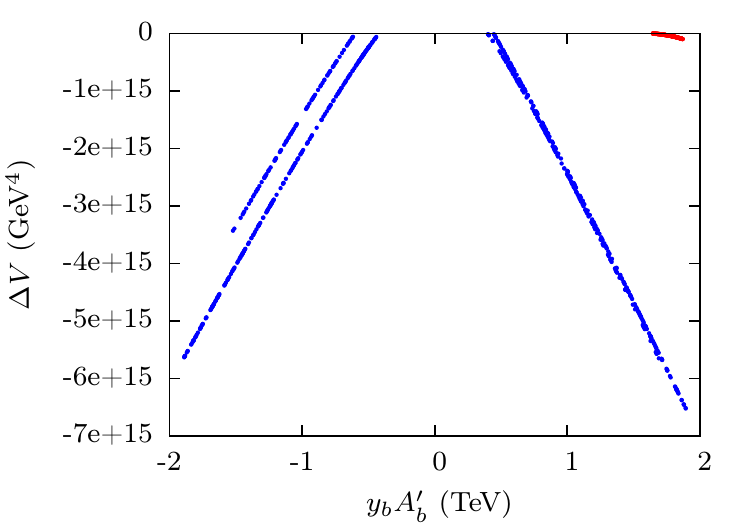}
		}
		\caption{\it $\Delta V$ represents the depth of the potential at the deeper CCB vacuum with respect the
			to the field origin. Fig.\ref{fig:pot1a} shows the variation of $\Delta V$  
			with $y_tA_t'$ for $\stop$ \Vevs~ scenario. 
			Fig.\ref{fig:pot1b} depicts the $y_bA_b'$ dependence
			of $\Delta V$ for $\sb$ \Vevs~ scenario.  The red and
			blue points correspond to $\tb=50$ and $\tb=5$ respectively. 
			Lowering of the potential depending on the suitable 
			values of $A_t'$ and $A_b'$ is evident from the above plots.  
			Large $|A_t'|$ and $|A_b'|$ thus essentially can lower the potential
			depending on the given choices of the other pMSSM parameters that are 
			kept fixed. }        
		\label{fig:pot1}
	\end{center}
\end{figure} 
Now we focus on the NH parameter space to identify the nature of
CCB vacuum in detail, by assigning \Vevs~ to $\stop$ and $\sb$ fields
separately. This is essential 
in order to identify the dangerous directions in the field space while
studying the role $A_t'$ and $A_b'$ in modifying the vacuum structure of the model. Since pNHSSM is principally motivated by naturalness,
we choose $\mu=200$ GeV for the entire analysis.     
Fig.\ref{fig:pot1} 
demonstrates the variation of $\Delta V$ with the
NH trilinear parameters for the vacuum deeper than the DSB one, 
where $\Delta V$ represents the depth of the 
scalar potential with respect to the origin. 
Fig.\ref{fig:pot1a} (\ref{fig:pot1b}) corresponds to the $\stop$ ($\sb$) \Vevs~ scenario. 
The red (blue) color refers to $\tb=50$ ($\tb=5$).  
We keep $\mtl$, $\mtr$ and $\mbr$ fixed at $2$ TeV. The MSSM trilinear term $y_t A_t$ ($y_b A_b$) is fixed at 2 TeV (0). The soft parameter $B_\mu$ in 
Eq.\ref{eq:scalrpot1} is kept fixed at 
$2\times 10^5$ GeV$^2$. The relevant NH parameters\footnote{In {\tt SPheno} convention,
the input trilinear soft SUSY breaking terms for sfermions are $y_f A_f$.} for the scenarios with $\stop$ and $\sb$ \Vevs~are given by the following ranges.
\begin{eqnarray}
\label{scanpotential}
-6~\mathrm{TeV} \leq y_tA_t' \leq 6~\mathrm{TeV} \notag\\  
-2~\mathrm{TeV} \leq y_bA_b' \leq 2~\mathrm{TeV}.
\end{eqnarray}
$A_t'$ and $A_b'$ clearly play a significant role in lowering the scalar potential. 
It is to be noted further, that the range of $y_b A_b'$ is kept small. This is
because large $A_b'$ leads to very large SUSY threshold corrections to bottom 
yukawa coupling and thus, the spectrum generator {\tt SPheno} aptly reports error. Recently, 
the effect of such corrections to $y_b$ has been studied in the context of vacuum stability in 
the MSSM \cite{Bobrowski:2014dla, Hollik:2015pra}.  
For the given choice of parameters, the depth of the DSB vacuum is many orders shallower than the depth at the deeper CCB vacuum. Furthermore, we observe from Fig.\ref{fig:pot1a} and \ref{fig:pot1b} that depth of the potential at the CCB vacuum is deeper for $\tb=5$ case.
Stop sector being the major contributor to the effective potential ($y_tA_t=2$ TeV gives large contribution
even for vanishingly small NH parameters), the effect gets enhanced for $\tb=5$ compared to $\tb=50$ due to large $y_t$ for the former case. We also observe from Fig.\ref{fig:pot1a} and \ref{fig:pot1b} that the depth of 
the potential at $\tb=5$ for the case with $\sb$ \Vev~ is a bit deeper compared to the
$\stop$ \Vev~one. This is due to the additional contribution arising from sbottom sector 
for large $A_b^\prime$ in the $\sb$ case.

It is to be noted in Fig.\ref{fig:pot1b} that there are two distinct blue lines for $A_b^\prime <0$. This is purely a numerical artifact due to the fact that the minimization routine {\tt Minuit} used in {\tt Vevacious} sometimes misses out some of the minima during random scan. Thus, in some cases, it reports a different CCB vacuum for similar
parameter sets. In the framework of {\tt Vevacious}, the \Vevs~ are bounded by
a hypercube of length $20Q$. We keep the renormalization scale $Q$ at a value
$\sqrt{\mtl \mtr}$ all through this work.
\begin{figure}[t]
	\begin{center}
		\subfigure[]{%
			\label{fig:pot2a}
			\includegraphics[width=0.4\textwidth]{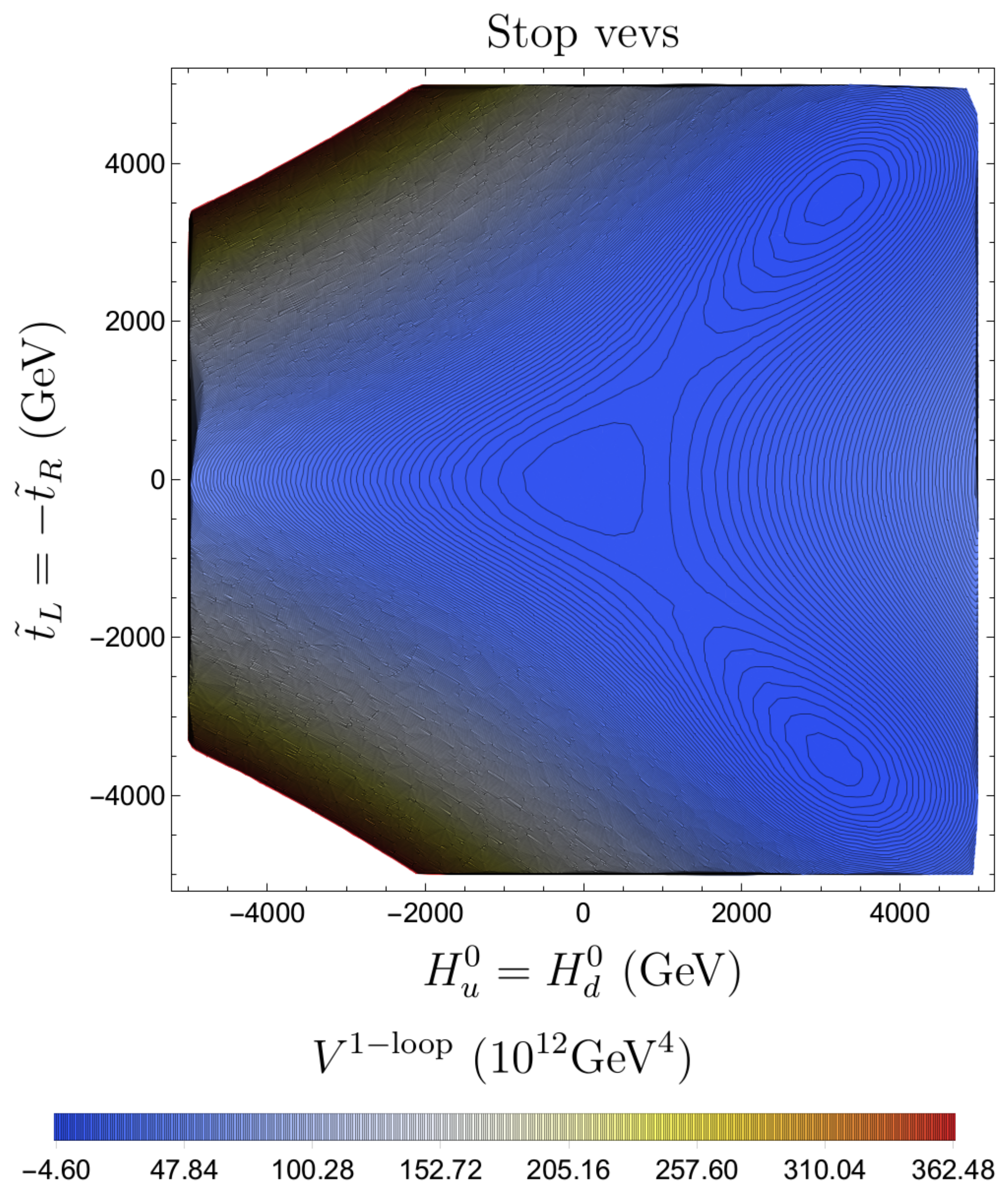}
		}%
		\hskip 0.2in
		\subfigure[]{%
			
			\label{fig:pot2b}
			\includegraphics[width=0.4\textwidth]{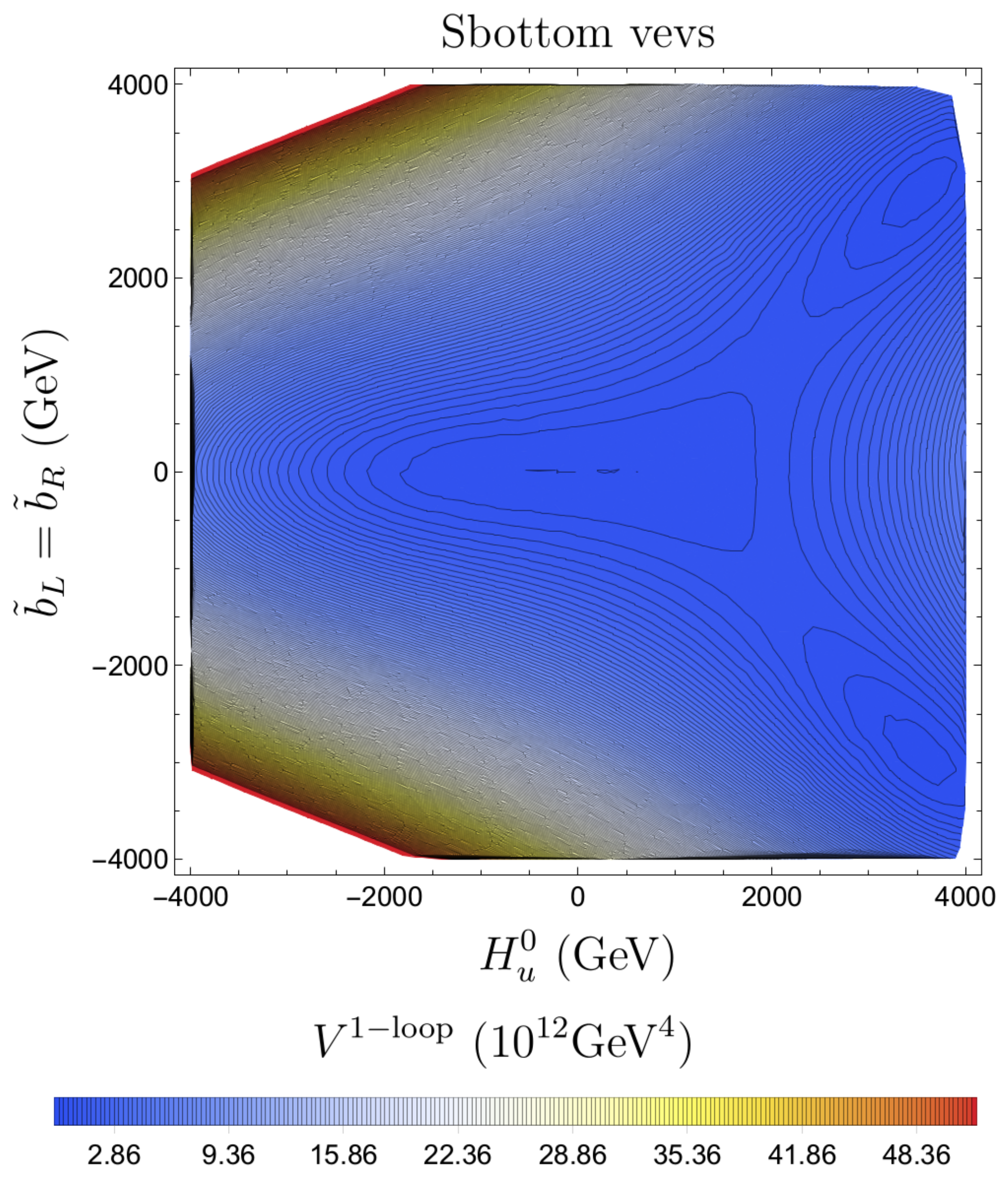}
		}
		\caption{\it The encircled closed contours surround the
			minima of the potential.
			The central contour encloses the DSB minima, while the other two
			contours that encircle regions with non-zero $\Vevs$ for stops (sbottoms) in Fig.\ref{fig:pot2a} (\ref{fig:pot2b}) represent the global CCB minima. }        
		\label{fig:pot2}
	\end{center}
\end{figure} 

In the Eqns. \ref{ccbconst} and \ref{ccbconsb} for the constraints on non-holomorphic parameters, all 
the \Vevs~ are considered to be equal. However, this is not a realistic 
scenario. In Fig.\ref{fig:pot2a} and Fig.\ref{fig:pot2b}, we present  schematic representation of contours for the cases for stop and sbottom \Vevs~ respectively. 
The fixed parameters are given as follows.
\begin{equation}
y_tA_t=2~\mathrm{TeV}, \mu=200~\mathrm{GeV},  \mtl=1~\mathrm{TeV}, \mtr=2~\mathrm{TeV},  \mbr=1~\mathrm{TeV}.
\end{equation}
While studying the effect of stop fields, we choose $y_t A_t'=-3600$ GeV with vanishing $A_b'$. On the other hand for the scenario with sbottom \Vevs~, $y_b A_b'$ is kept at $800$ GeV with vanishing $A_t'$. The encircled closed contours surround the minima of the potential.
The central contour encloses the DSB minima, while the other two
contours that encircle regions with non-zero $\Vevs$ for stops (sbottoms) represent  CCB minima. For the stop \Vevs~  (Fig.\ref{fig:pot2a}), we choose $H_u^0=H_d^0$  direction and for the sbottom \Vev~  (Fig.\ref{fig:pot2b}), we choose $H_d^0 =0$ direction in order to capture the global CCB minima and the DSB vacuum.
In the subsequent sub-sections, we analyze the CCB vacua due to stop fields, sbottom fields and finally, 
combination of both stop and sbottom fields on a  broader region of parameter space compared to the
one presented above.
\subsection{CCB minima associated with stop fields}
\begin{figure}[ht]
	\begin{center}
		\subfigure[]{%
			\label{fig:xtpmh10}
			\includegraphics[width=0.51\textwidth]{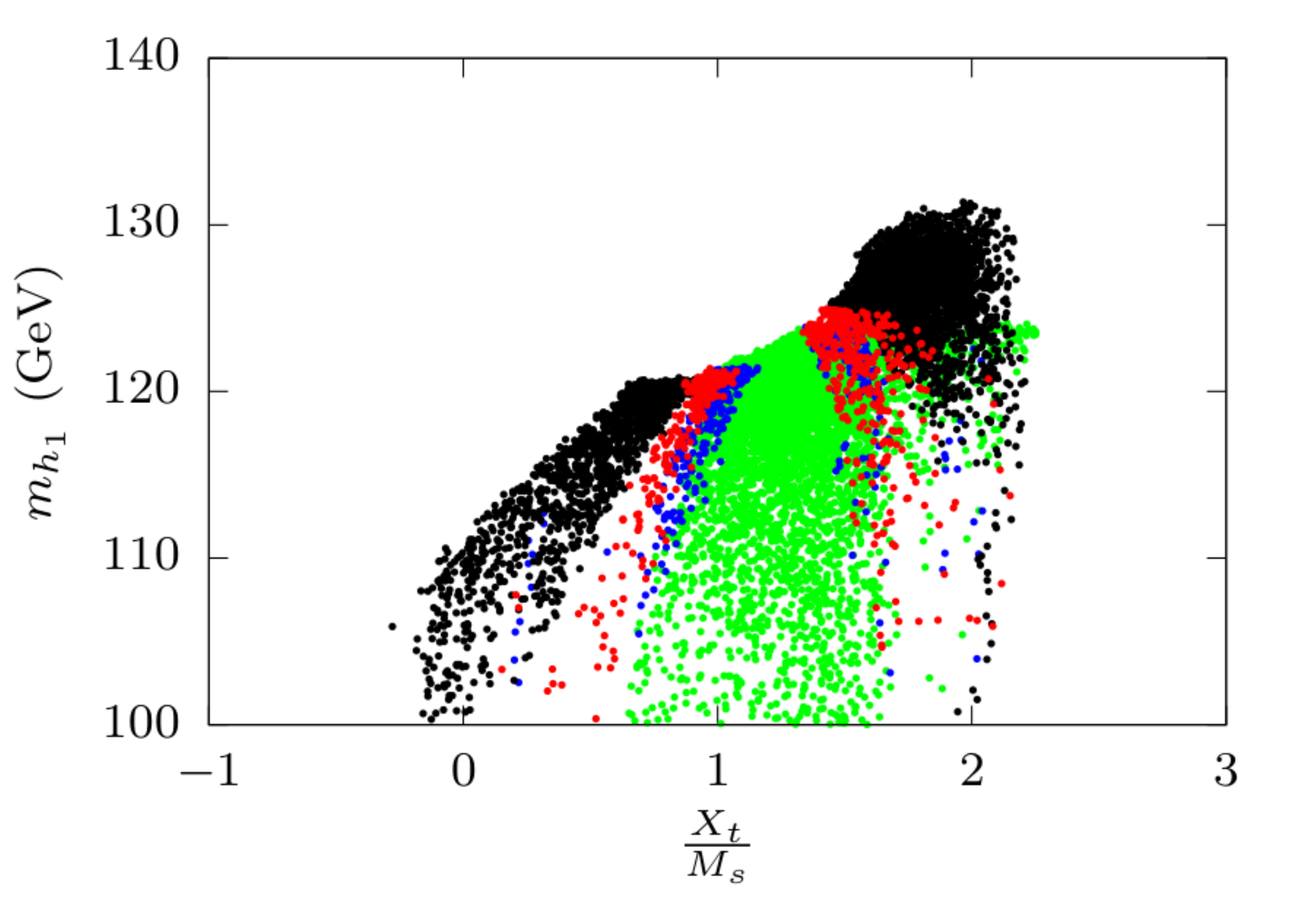}
		}%
		\subfigure[]{%
			
			\label{fig:xtpmh50}
			\includegraphics[width=0.51\textwidth]{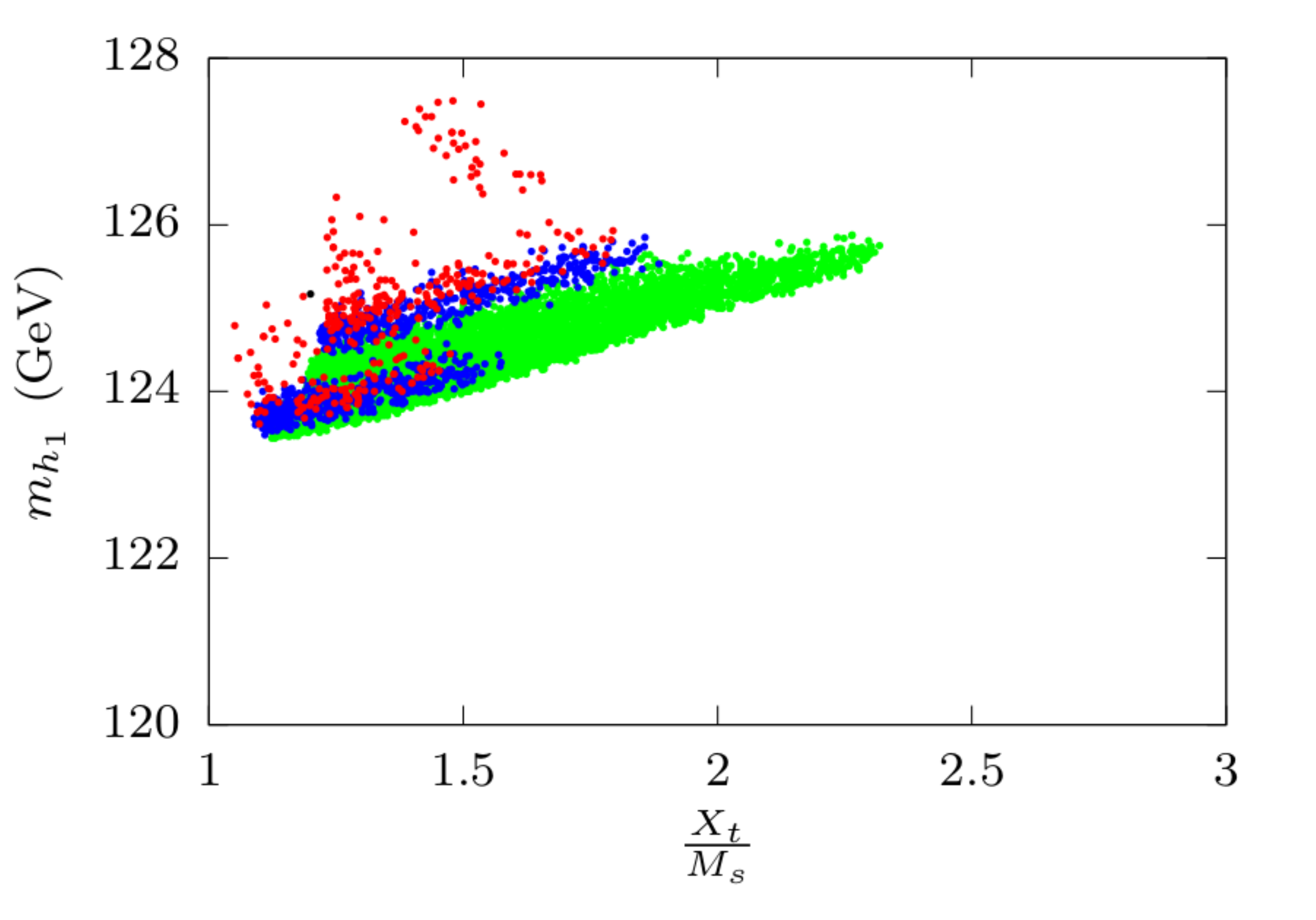}
		}
		\caption{\it The vacuum stability profile is plotted in $\frac{X_t}{M_s}-m_{h_1}$ plane, where $X_t=A_t-(\mu+A_t^\prime)\cot\beta$ and
		$M_s=\sqrt{m_{\stopone} m_{\stoptwo}}$ with tree level stop masses. The left (right) plot corresponds to $5\leq \tb \leq 10$ $(40\leq \tb \leq 50)$. Green and blue regions correspond to stable and long-lived vacuum respectively. The red (black) points refer to thermally excluded (quantum mechanically unstable) vacuum.}        
		\label{fig:xtpmh}
	\end{center}
\end{figure} 
Keeping an eye on Eq.\ref{ccbconst} we define $\mathcal{M}_{\stop}^2 = 
\left[ m_1^2 + m_2^2 + m_{\stopl}^2 + 
m_{\stopr}^2  -2B_{\mu}\right]$ and $\widetilde{A_t}^2= \left(|A_t| +|\mu| + |A_t'|\right )^2$. 
These variables will play a crucial role in determining the 
applicability of the analytic constraint of Eq.\ref{ccbconst}.
We also define similar variables $\mathcal{M}_{\sb}^2=( m_1^2 + m_2^2 + m_{\stopl}^2 + 
m_{\sbottomr}^2  -2B_{\mu})$ and $\widetilde{A_b}^2= \left(|A_b| +|\mu| + |A_b'|\right )^2$ to be used while studying the CCB constraints arising
out of non-vanishing \Vevs~ of $\sb$ fields.

We probe the vacuum stability of the model for both smaller 
$(5\leq \tb \leq 10)$ and larger values  $(40\leq \tb \leq 50)$ of $\tb$. 
We analyze in a minimal four-\Vev~ set-up considering non-zero \Vevs~ for 
the Higgs and the stops.  We vary the relevant parameters in the following ranges.
\begin{eqnarray}
\label{scanonlystop}
-1~\mathrm{TeV} \leq \mu' \leq 1~\mathrm{TeV}  \notag \\ 
-6~\mathrm{TeV} \leq y_t A_t' \leq 6~\mathrm{TeV}  \notag \\
500~\mathrm{GeV} \leq \mtl \leq 2~\mathrm{TeV} . 
\end{eqnarray}

We keep $y_tA_t$ fixed at $2$ TeV, and set all other holomorphic 
soft trilinear parameters to zero. All scalar mass parameters 
excluding $\mtl$ are fixed at $2$ TeV.  Since we study the vacuum 
structure with specific motive of investigating the role of NH 
parameters, we do not scan over holomorphic trilinear parameters. 
Other fixed parameters are same as that of Fig.\ref{fig:pot1a}.
Fig.\ref{fig:xtpmh} shows the vacuum stability profile
in $X_t-m_{h_1}$ plane.
\begin{figure}[t]
\begin{center}
\subfigure[]{%
\label{fig:atpmh10}
\includegraphics[width=0.48\textwidth]{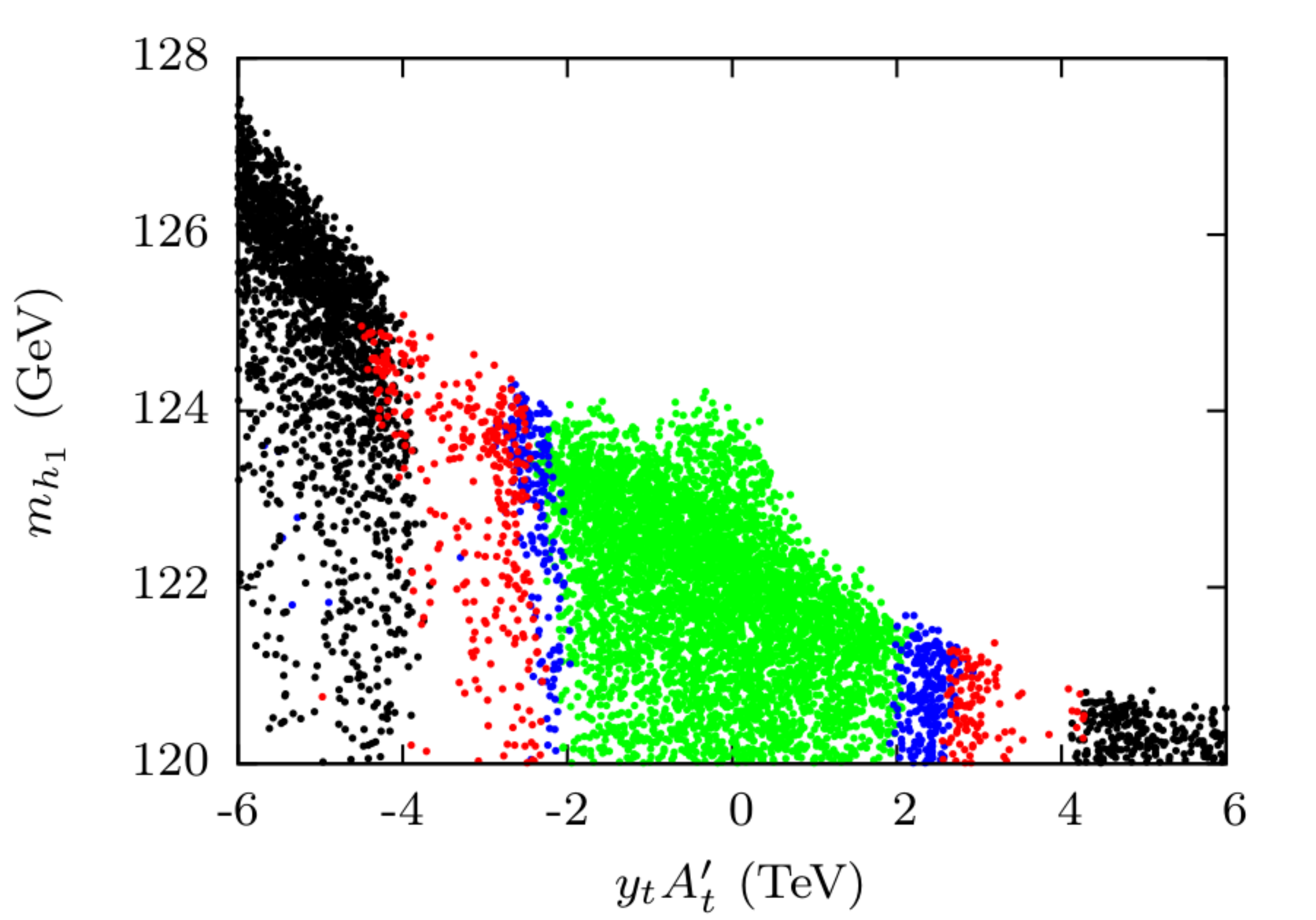}
}%
\subfigure[]{%
\label{fig:atpmh50}
\includegraphics[width=0.48\textwidth]{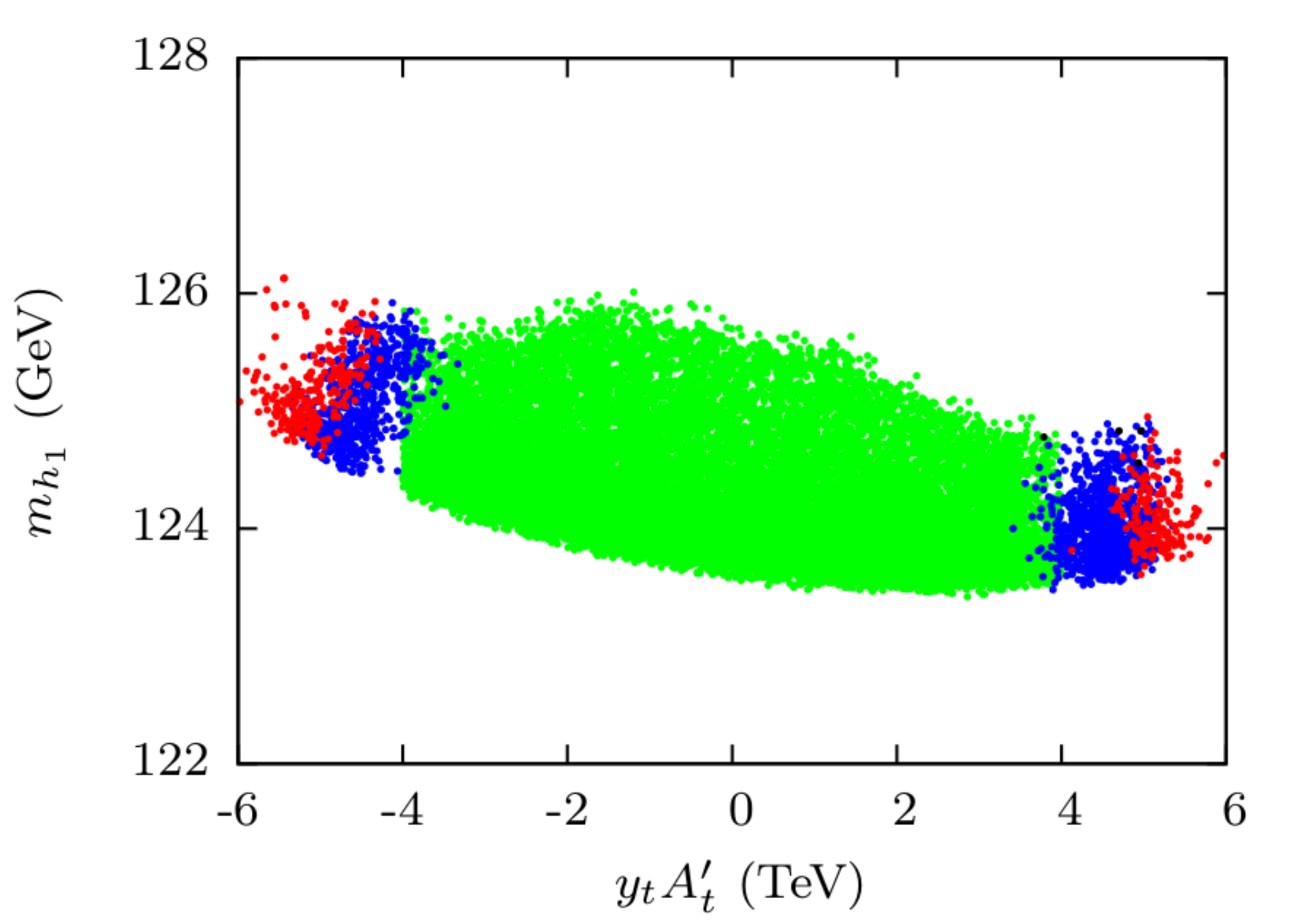}
}
\caption{\it The stability profile of DSB 
minima in $y_t A_t' -m_{h_1}$ plane corresponding to the scan of 
Eq.\ref{scanonlystop}. Figs.\ref{fig:atpmh10} and \ref{fig:atpmh50} 
corresponds to $(5 < \tb < 10)$ and $(40 < \tb < 50)$ respectively. The 
color codes are same as that for Fig.\ref{fig:xtpmh}.}        
\label{fig:atpmh}
\end{center}
\end{figure}
In spite of having no influence on vacuum stability, 
$\mu'$ is scanned over a moderate range. This is due to the fact 
that it contributes to $m_{h_1}$ via chargino loop \cite{Ibrahim:2000qj,Haber:1990aw}. 
The green and blue points are associated 
with stable and long-lived DSB minima. They are collectively 
referred to as safe vacua.  The black and the red colored 
regions are excluded since the DSB minima associated with them are 
rendered unstable with respect to quantum tunneling to deeper CCB minima  
and thermal effects respectively.
Fig.\ref{fig:xtpmh10} and \ref{fig:xtpmh50} correspond 
to low and high $\tb$ regimes. 
In Fig.\ref{fig:xtpmh10}, we see stable points (green) cluster 
around $\frac{X_t}{M_s} \approx 1$ with close by metastable points. This is mainly because of the
choice of $y_t A_t=2$ TeV with $\mu=200$ GeV. Thus, most of the points appear on positive
side of $\frac{X_t}{M_s}$. The vacuum stability profile for the MSSM \cite{Camargo-Molina:2014pwa} in 
$X_t-m_{h_1}$ plane is qualitatively very much similar to Fig.\ref{fig:xtpmh10}. Fig.\ref{fig:xtpmh50} appear
to be a bit different from Fig.\ref{fig:xtpmh10}. This is because mass of the Higgs boson gets 
boosted for chosen value of $y_tA_t$ (2 TeV) and $\tb=50$. There are no quantum mechanically short-lived points (black) 
for Fig.\ref{fig:xtpmh50}, whereas there are large number of black points in Fig.\ref{fig:xtpmh10}.
\begin{figure}[t]
\begin{center}
\subfigure[]{%
\label{fig:atpmst10}
\includegraphics[width=0.48\textwidth]{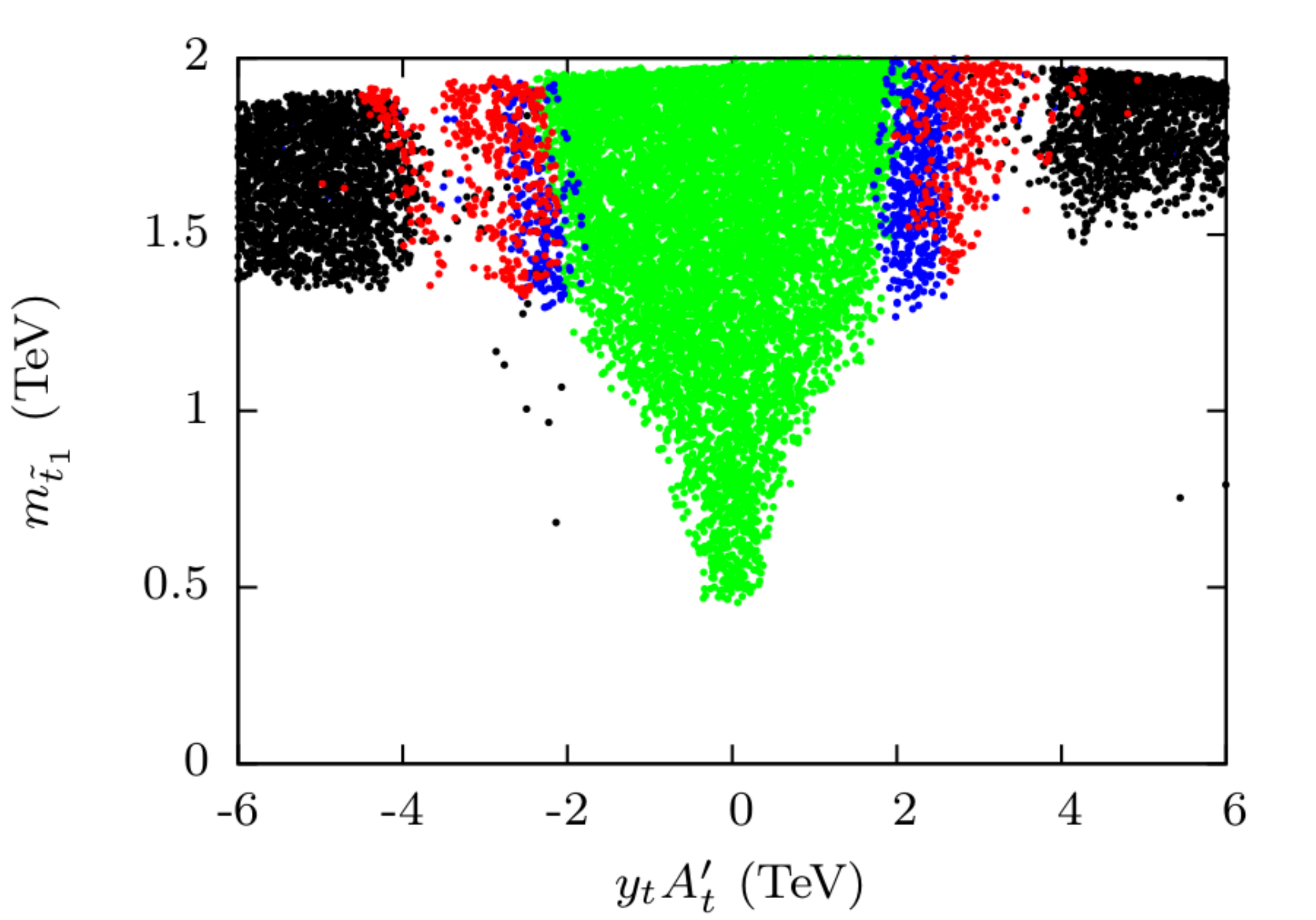}
}%
\subfigure[]{%
\label{fig:atpmst50}
\includegraphics[width=0.48\textwidth]{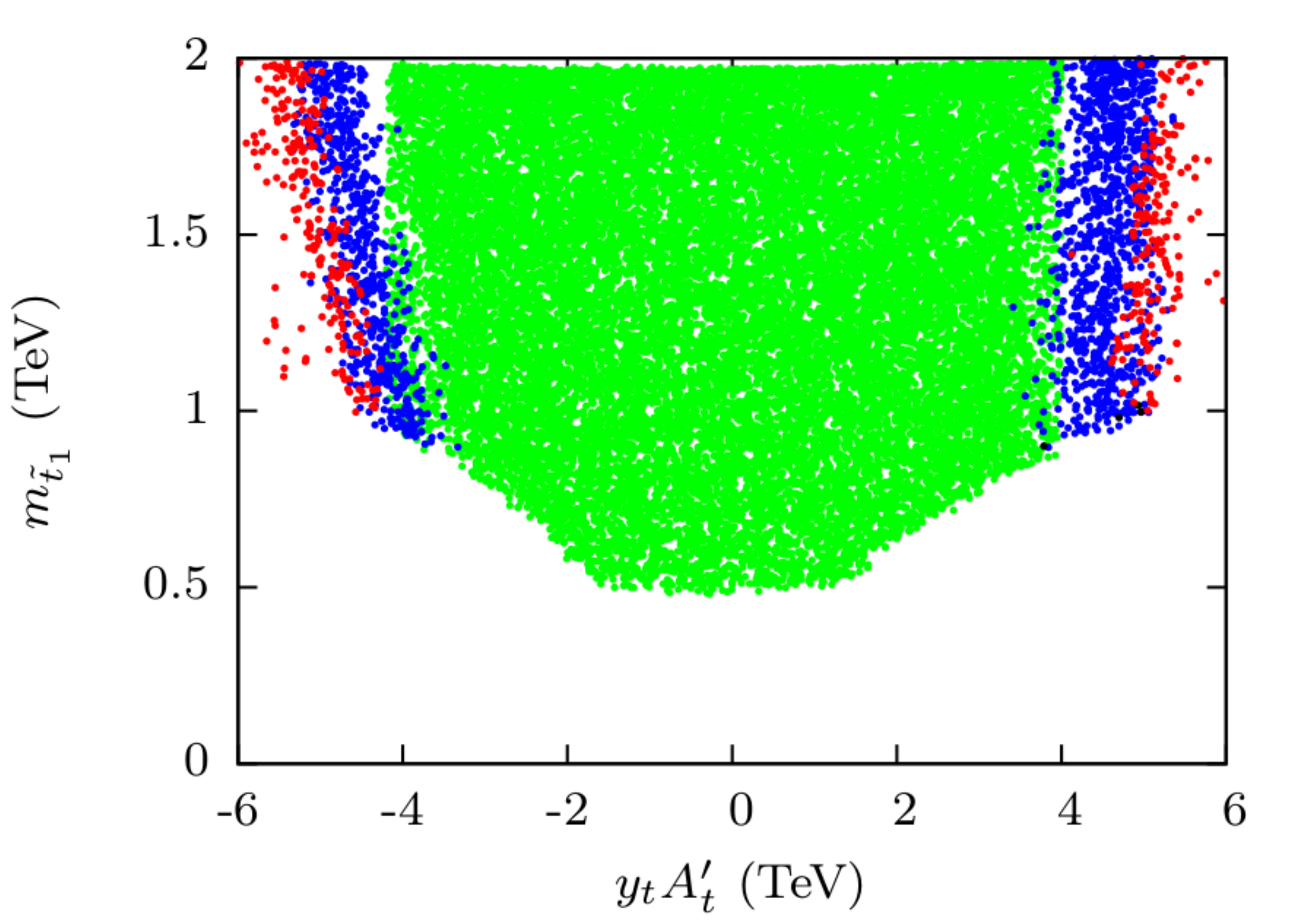}
}
\caption{\it The stability profile of the DSB 
minima in $y_tA_t'-m_{\stopone}$ plane corresponding to the scan of 
Eq.\ref{scanonlystop}. Figs.\ref{fig:atpmst10} and \ref{fig:atpmst50} 
corresponds to $(5 < \tb < 10)$ and $(40 < \tb < 50)$ respectively.The 
color codes are same as that for Fig.\ref{fig:xtpmh}.}        
\label{fig:atpmst}
\end{center}
\end{figure}

Fig.\ref{fig:atpmh} shows the dependence of $m_{h_1}$ on the 
$y_tA_t'$ for $\tb$ scanned over 
two different domains.  
Fig.\ref{fig:atpmh10} shows the stability profile for 
$5\leq \tb \leq 10$. 
The spread of $m_{h_1}$ for particular value 
of $y_tA_t'$ is due to the effect of variation of $\mtl$ and $\tb$ 
on the radiative corrections arising out of stop loops (Eq. \ref{stop_loop}). 
Fig.\ref{fig:atpmh50} shows the similar plot 
for $40\leq \tb \leq 50$.  As discussed in Sec.\ref{NHSSM}, $A_t'$
contribution in the radiative correction to $m_{h_1}$ via $\stop$ 
loop is $\tb$ suppressed. As a result the effect of $y_tA_t'$ on $m_{h_1}$ 
is more prominent in low $\tb$ region. Thus, for Fig.\ref{fig:atpmh50} 
we see that the variation of $m_{h_1}$ is significantly small compared 
to Fig.\ref{fig:atpmh10}. In both the plots, the central 
region characterized by comparatively lower $|y_tA_t'|$ is associated 
with stable DSB vacua, whereas large $|y_tA_t'|$ region is associated 
with unsafe DSB minima. In between the two regions, there exists a small 
zone associated with long-lived states (blue).
We find that for low $\tb$ and $|y_t A_t'| \gsim 3$ TeV the DSB minima 
becomes unsafe and they are excluded via quantum tunneling or thermal effects. 
On the contrary, for $40 \leq \tb \leq 50$, comparatively large values 
of $|y_t A_t'|$  ($ < 4$ TeV), is allowed in the region of parameter 
space associated with safe vacuum. This is due to the fact that the term 
in the potential associated with $A_t'$ is suppressed for large $\tb$.  

Fig.\ref{fig:atpmst} shows the vacuum stability profile in $y_tA_t'-m_{\stopone}$ 
plane for same set of scans. As before, Fig.\ref{fig:atpmst10} (Fig.\ref{fig:atpmst50})
represents the case with $5 \leq \tb \leq 10$ ($40 \leq \tb \leq 50$). 
We observe from both the plots that $m_{\stopone}$ 
could be rather large in the region of metastability triggered by large $|y_tA_t'|$.
We have kept $m_{\tilde{t}_R}$ fixed at 2 TeV. Hence large $m_{\tilde{t}_L}$ along with
large $|y_t A_t'|$ facilitate the appearance of deeper CCB vacua by inducing large 
mixing between ${\tilde{t}_L}$ and ${\tilde{t}_R}$. Thus, $m_{\tilde{t}_1}$ becomes
 large for large $|y_tA_t'|$. We also 
notice that due to $\tb$ suppression, the spread of green points in 
Fig.\ref{fig:atpmst50} is more compared to Fig.\ref{fig:atpmst10}.
Furthermore, the flat green edge ($m_{\tilde{t}_L} \approx m_{\tilde{t}_1}\approx 500$ GeV) is also broader 
compared to Fig.\ref{fig:atpmst10} for the same reason.

\begin{figure}[t]
\begin{center}
\subfigure[]{%
\label{fig:consmh10}
\includegraphics[width=0.51\textwidth]{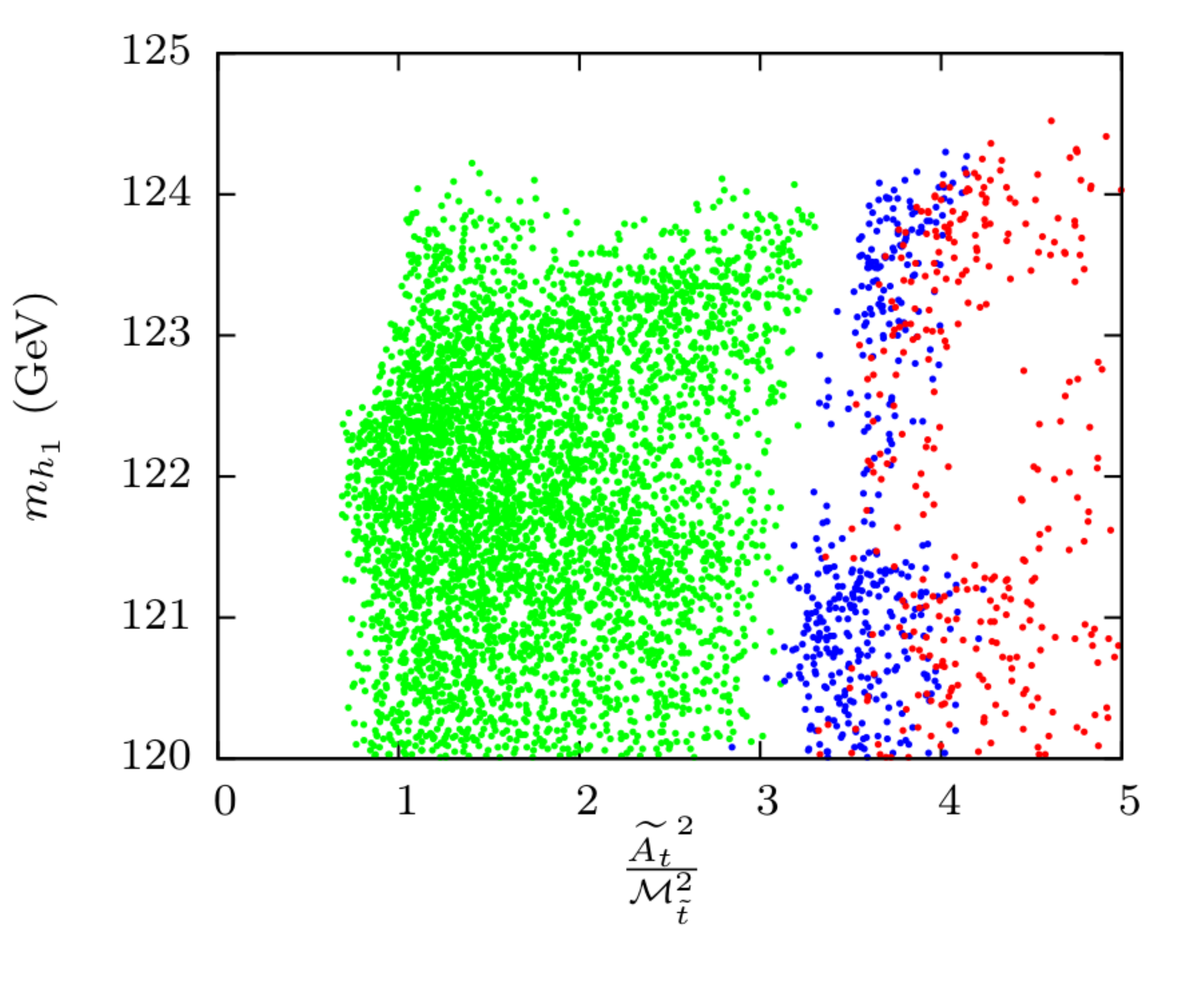}
}%
\subfigure[]{%
\label{fig:consmh50}
\includegraphics[width=0.51\textwidth]{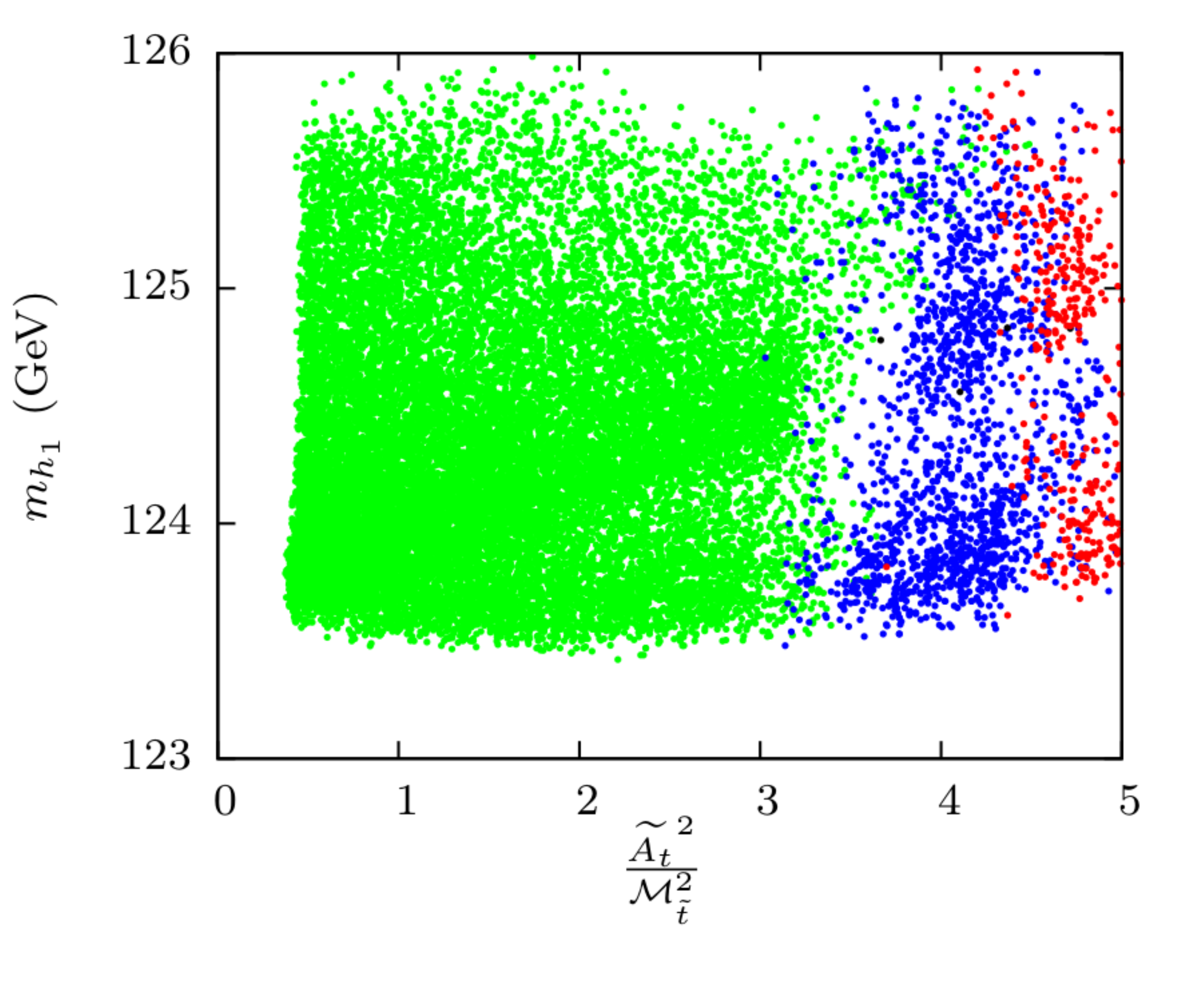}
}
\caption{\it The stability profile of the DSB vacuum in $\frac{\widetilde{A_t}^2}{\mathcal{M}_{\stop}^2}-m_{h_1}$ plane. Fig.\ref{fig:consmh10} (\ref{fig:consmh50}) refers to $5\leq \tb \leq 10$ ($40\leq \tb \leq 50$). 
The color codes are same as that for fig.\ref{fig:xtpmh}.} 
\label{fig:consmh}
\end{center}
\end{figure}     
Fig.\ref{fig:consmh} shows the stability profile with $m_{h_1}$ along
$y$-axis and the relevant factor 
$\frac{\widetilde{A_t}^2}{\mathcal{M}_{\stop}^2}$ (see Eq.\ref{ccbconst}) along the $x$-axis.
Here, we explore the applicability 
of the analytic CCB constraints in the NHSSM, exclusively in 
four \Vevs~ scenario with non-vanishing \Vevs~ for stops and 
Higgs fields. The variable in the horizontal axis is predicted 
to be less than $3$ for stable DSB minima according to Eq. 
\ref{ccbconst}. It appears that such constraint holds
quite reliably for the chosen region of parameter space. However, since the analytic constraint of Eq. \ref{ccbconst} was derived under the consideration of stable vacua only, the constraint may be relaxed in the present scenario where, we also include long-lived DSB minima as viable vacua of the theory. This is evident from the extent of the long-lived states in the region where $3 \lsim \frac{\widetilde{A_t}^2}{\mathcal{M}_{\stop}^2} \lsim 4$.  We observe that the safe vacua spread over a wider range of $\frac{\widetilde{A_t}^2}{\mathcal{M}_{\stop}^2}$ for large $\tb$. This is quite expected as the contribution of $ A_t'$ is $\tb$ suppressed. Moreover for large $\tb$, $y_t$ being comparatively smaller, the rate of quantum tunneling is decreased enhancing the presence of long-lived vacua over a wider range of $y_tA_t'$ and $\frac{\widetilde{A_t}^2}{\mathcal{M}_{\stop}^2}$.  
In the next section, a study of the deeper CCB vacua arising from the 
$\Vevs$ of the sbottoms is presented. 

\subsection{CCB minima associated with sbottom fields}
As discussed in Sec.\ref{NHSSM} the global CCB minima associated 
with $\stop$ fields are more dangerous with respect to 
the tunneling rate due to large yukawa coupling $(y_t)$.  
The $\sb$  fields may also become important particularly 
for large $\tb$. Hence, in this section, we  study the role 
of $y_b A_b'$ associated with $\sbottoml$ and 
$\sbottomr$ in determining the fate of DSB minima. 
We probe the vacuum stability of the model for both 
smaller  $(5\leq \tb \leq 10)$ and 
larger values  $(40\leq \tb \leq 50)$ of $\tb$.  
Considering the effects on vacuum stability, we 
vary the relevant parameters in the following ranges.
\begin{eqnarray}
\label{scanonlysbot}
-1~\mathrm{TeV} \leq \mu' \leq 1~\mathrm{TeV}  \notag \\ 
-2~\mathrm{TeV} \leq y_bA_b' \leq 2~\mathrm{TeV}  \notag \\
500~\mathrm{GeV} \leq \mtl \leq 2~\mathrm{TeV}   \notag \\
500~\mathrm{GeV} \leq \mbr \leq 2~\mathrm{TeV} 
\end{eqnarray}
We keep $y_tA_t$ fixed at $2$ TeV, and set all other holomorphic soft 
trilinear parameters to zero. All scalar mass parameters excluding 
$\mtl$ and $\mbr$ are fixed at $2$ TeV. 
\begin{figure}[t]
\begin{center}
\subfigure[]{%
\label{fig:sbot10}
\includegraphics[width=0.51\textwidth]{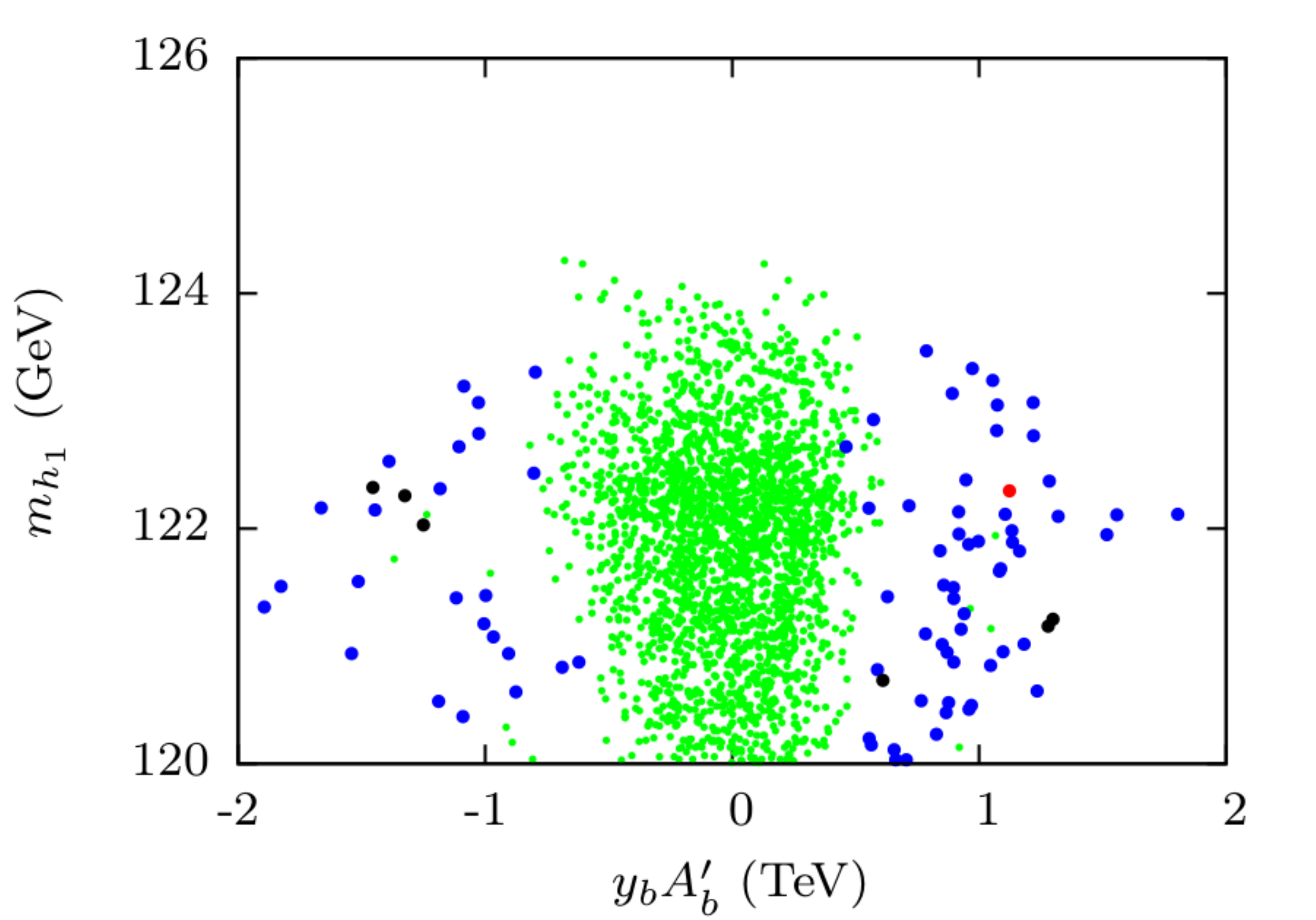}
}%
\subfigure[]{%
\label{fig:sbot40}
\includegraphics[width=0.51\textwidth]{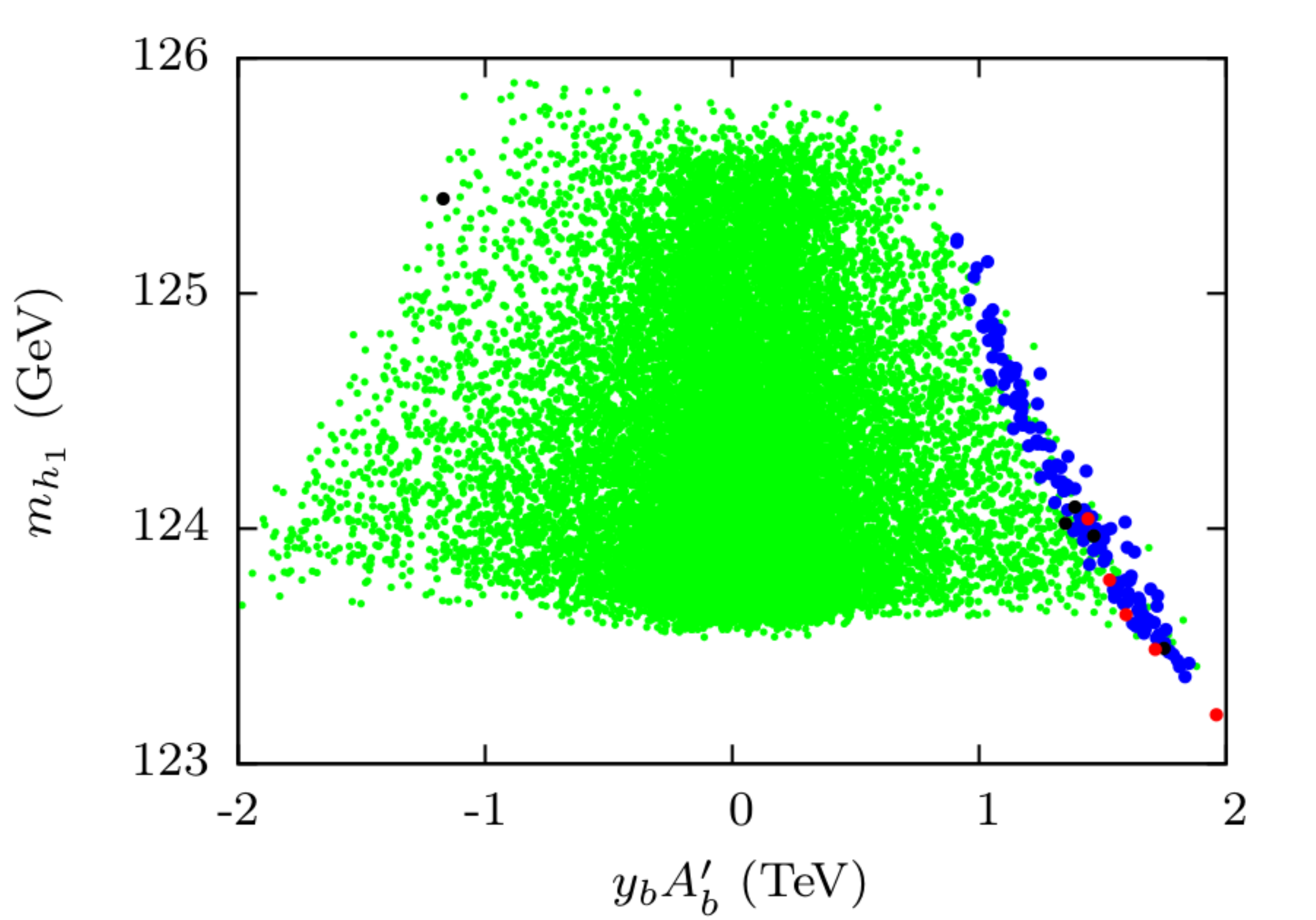}
}
\caption{\it Fig.\ref{fig:sbot10} shows the stability profile in 
$y_b A_b'-m_{h_1}$ plane for $5 < \tb <  10$ corresponding to scan of Eq.\ref{scanonlysbot}. Fig.\ref{fig:sbot40} is a
similar plot for $ 40 < \tb < 50$. The color codes for stability 
profile is same as that for Fig.\ref{fig:xtpmh}.}        
\label{fig:sbot}
\end{center}
\end{figure}
Fig.\ref{fig:sbot10} shows the stability profile in $y_b A_b'- m_{h_1}$
plane for low $\tb$ $(5 < \tb < 10)$. As expected, we see that 
the DSB minima are safe for most of the values of $A_b'$. 
This is consistent with the discussion in Sec.\ref{NHSSM}, as for small 
$\tb$ the NH terms associated with $\sb$ are less dominant.
It appears that most of the metastable points are actually long-lived 
and thus, these are considered safe. For low $\tb$, $y_b$ is very small
and mixing between $\sbottoml$ and $\sbottomr$ would only be enhanced 
if difference between $m_{\stopl}$ and $m_{\sbottomr}$ is rather small.
Hence the long-lived (blue) points have small difference between
$m_{\stopl}$ and $m_{\sbottomr}$. 
  
Fig.\ref{fig:sbot40} shows the stability profile
for large $\tb$. Here we see that large values of $y_bA_b'>0$ have
deeper CCB vacua compared to the corresponding DSB ones. The region $y_b A_b'<0$
hardly appear to possess metastable points. This is because the SUSY
threshold corrections to $y_b$ for the chosen value of $\mu=200$ GeV increases 
$y_b$ very significantly for $A_b'<0$ compared to $A_b'>0$. 
Thus, a particle mass spectrum calculated by {\tt SPheno}
contains a DSB vacuum consistent at two-loop level. The model parameters
deduced at this setup may not necessarily yield a similar DSB vacuum
when one-loop correction to scalar potential is employed. 
This makes {\tt Vevacious} reassign the input vacuum
to a rolled down vacuum configuration other than the DSB one determined
from the spectrum file. In our analysis, we are not considering such kind
of particle spectra where two-loop corrections change the structure of the potential
very significantly. This results in hardly any metastable points for $A_b'<0$. 
\begin{figure}[t]
\begin{center}
\subfigure[]{%
\label{fig:sbot210}
\includegraphics[width=0.51\textwidth]{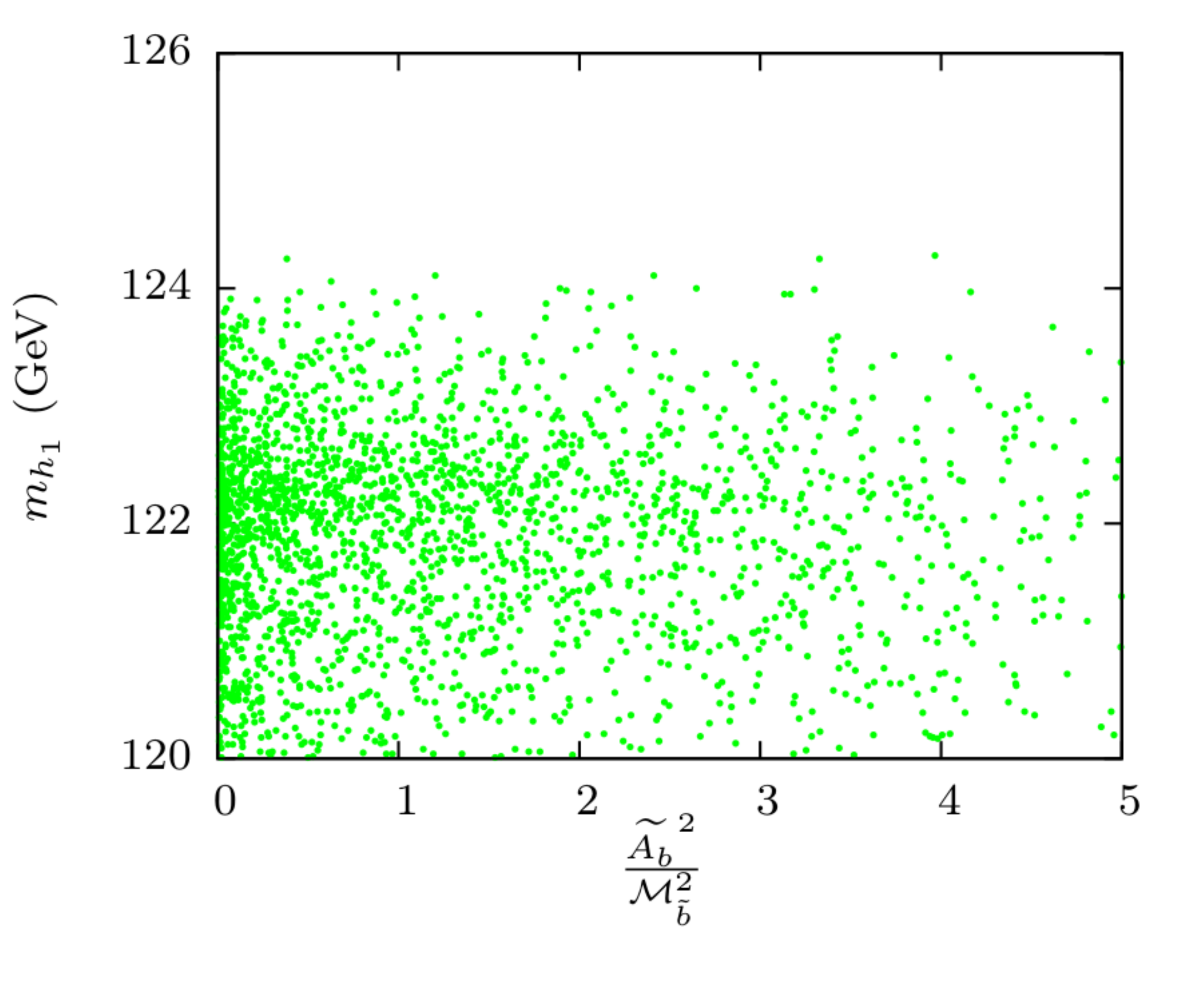}
}%
\subfigure[]{%
\label{fig:sbot240}
\includegraphics[width=0.51\textwidth]{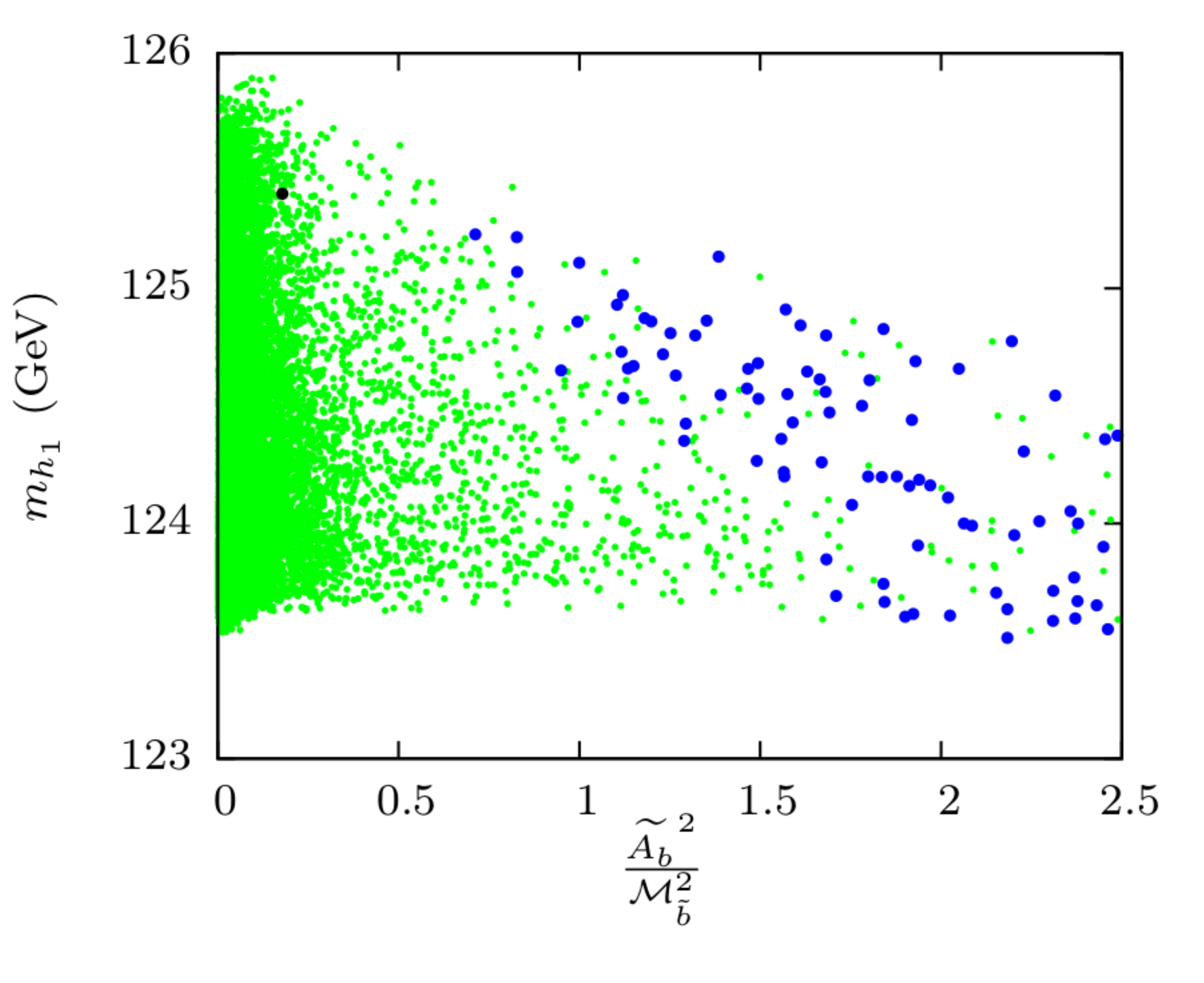}
}%
\caption{\it \it Fig.\ref{fig:sbot210} shows the stability 
profile in $\frac{\widetilde{A_b}^2}{\mathcal{M}_{\sb}^2}-m_{h_1} $ plane 
for $5 < \tb <  10$.  Fig.\ref{fig:sbot240} is similar plot for 
$ 40 < \tb < 50$. The color codes for stability profile is same 
as that for Fig.\ref{fig:atpmh}. It is evident from the plots 
that the analytic expression regarding CCB constraint
Eq.\ref{ccbconsb} is unimportant for small $\tb$, unlike the case of
large $\tb$ (see text for details).}
\label{fig:sbot2}
\end{center}
\end{figure}    
We further notice that $m_{h_1}$ spreads over a larger range for large $\tb$. 
This is due to the fact that the contribution of $|A_b'|$ to $m_{h_1}$ is enhanced by $\tb$ 
(Eq.\ref{sbottomeqn}).

In Fig.\ref{fig:sbot2}, we plot the stability profile in 
$\frac{\widetilde{A_b}^2}{\mathcal{M}_{\sb}^2}-m_{h_1}$ plane.
Fig.\ref{fig:sbot210} (Fig.\ref{fig:sbot240}) corresponds to small 
(large) $\tb$. For both $\tb$ regimes, the analytic constraint seems
to be unnecessary since almost all of the region of parameter space that 
was scanned over, corresponds to stable or long-lived vacua.
For large $\tb$, stable vacuum  
corresponds to region with rather smaller values of $\frac{\widetilde{A_b}^2}{\mathcal{M}_{\sb}^2}$
compared to the case with smaller $\tb$.
For smaller $\tb$, the metastable points appearing in Fig.\ref{fig:sbot10} do not appear for $\frac{\widetilde{A_b}^2}{\mathcal{M}_{\sb}^2}<5$. On the other hand, the metastable
points appearing in Fig.\ref{fig:sbot40} appear for smaller values of 
$\frac{\widetilde{A_b}^2}{\mathcal{M}_{\sb}^2}$. This because for smaller $\tb$, $y_b$ is
very small compared to the case with larger $\tb$. This renders $A_b'$ very large for 
smaller $\tb$ for a particular value of $y_b A_b'$ compared to larger $\tb$. Considering absolute stable vacua, we observe that metastability appears even when $\frac{\widetilde{A_b}^2}{\mathcal{M}_{\sb}^2}$ is significantly less than $3$ on contrary to what predicted by Eq.\ref{ccbconsb}. This is due to the wide variation of $y_b$ in the large $\tb$ regime. Hence $\frac{g_1^2+g_2^2}{24y_b^2}$ is not always negligible with respect to $1$. Thus, the factor $3$ does not arise. 
\subsection{CCB minima associated with both stop and sbottom fields}
In this section we study a more involved scenario characterized by the non-vanishing 
\Vevs~ for all the third generation squark fields along with  Higgs fields. 
We vary the relevant parameters in the following ranges.
\begin{figure}[t]
\begin{center}
\subfigure[]{%
\label{fig:sbotstop110}
\includegraphics[width=0.51\textwidth]{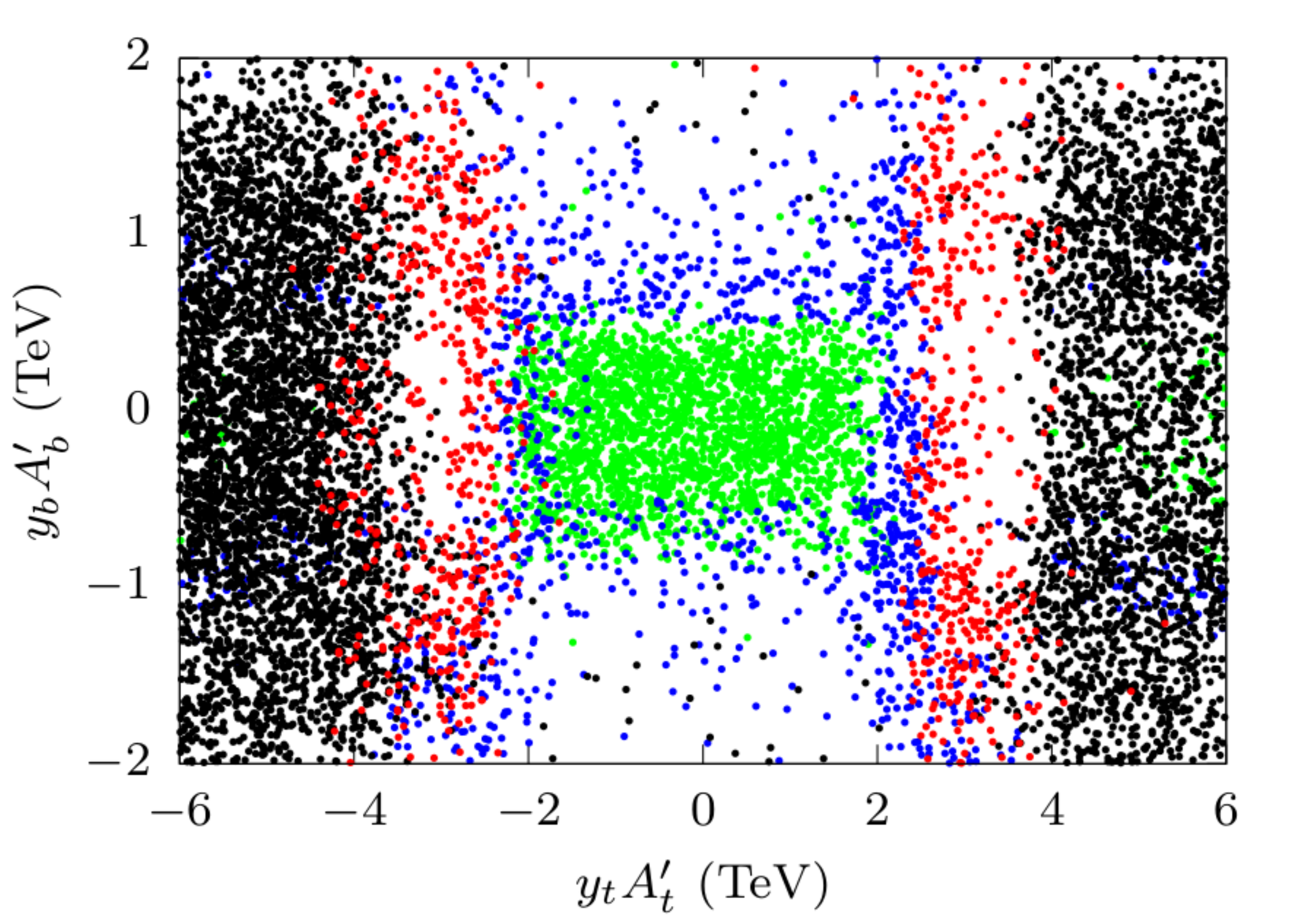}
}%
\subfigure[]{%
\label{fig:sbotstop140}
\includegraphics[width=0.51\textwidth]{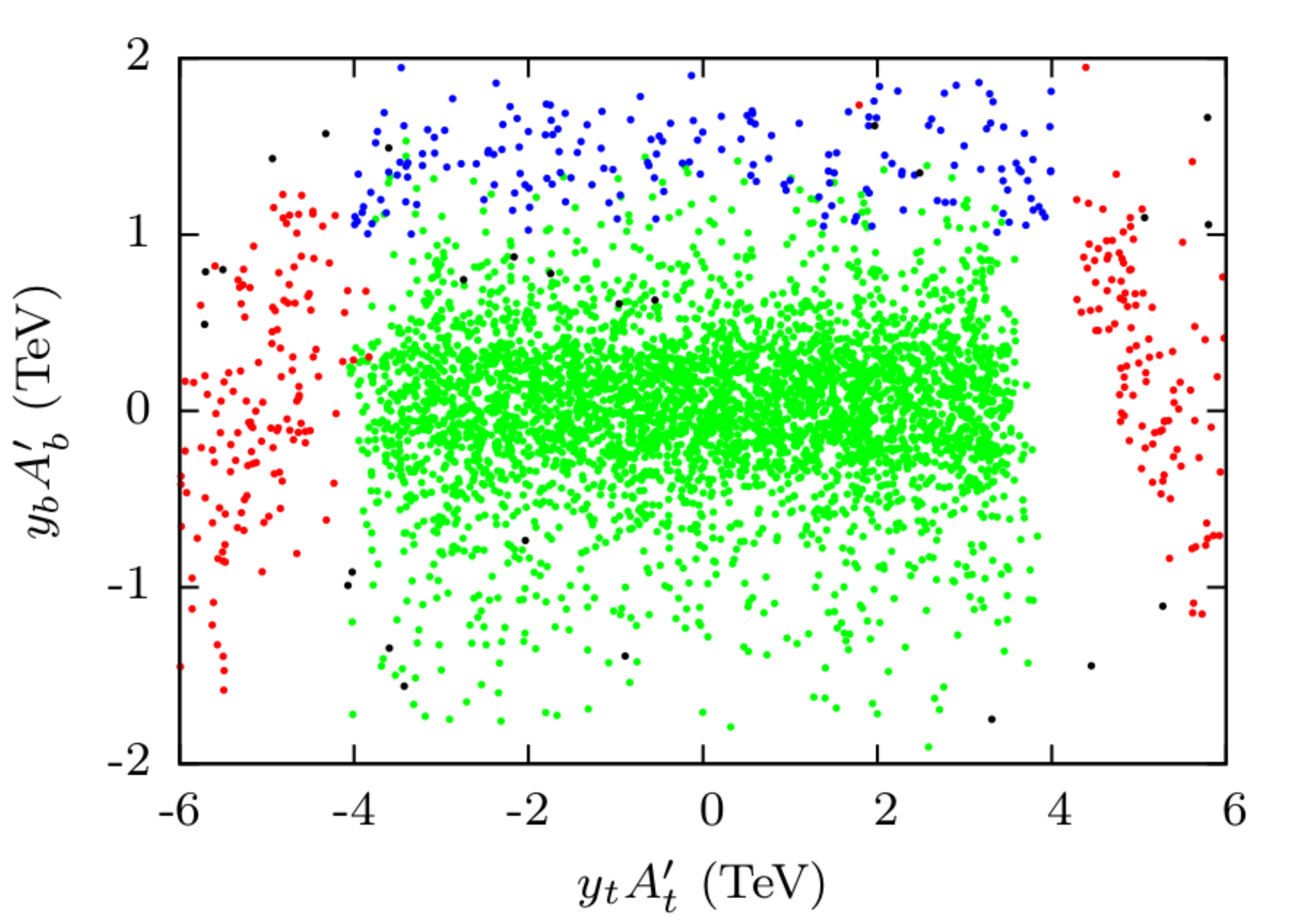}
}%
\caption{\it Figs.\ref{fig:sbotstop110} and \ref{fig:sbotstop140} classify 
by the parameter points according to the fate of DSB minima associated with 
them, in the $y_t A_t'-y_bA_b'$ plane, for low and high regime of $\tb$. Fig.\ref{fig:sbotstop1} 
corresponds to the scan of Eq.\ref{scanbothstopsbot}. It is 
evident that the effect of $A_t'$ is more prominent for low $\tb$, whereas 
$A_b'$ plays a crucial role for large $\tb$. }        
\label{fig:sbotstop1}
\end{center}
\end{figure}
\begin{eqnarray}
\label{scanbothstopsbot}
-1 ~\mathrm{TeV} \leq \mu' \leq 1~\mathrm{TeV},  \notag \\ 
-6 ~\mathrm{TeV} \leq y_tA_t' \leq 6~\mathrm{TeV},  \notag \\
-2~\mathrm{TeV} \leq y_bA_b' \leq 2~\mathrm{TeV},          \\
500~\mathrm{GeV} \leq \mtl \leq 2~\mathrm{TeV},   \notag \\
500~\mathrm{GeV} \leq \mbr \leq 2~\mathrm{TeV}.   \notag 
\end{eqnarray}
We keep $y_tA_t$ fixed at $2$ TeV and $\mu$ at $200$~GeV. All sfermion 
mass parameters except $\mtl$ and $\mbr$, are fixed at $2$ TeV.
Fig.\ref{fig:sbotstop1} shows the stability profile in 
the $y_t A_t'-y_b A_b'$ plane. The color codes for stability profile are same 
as that for Fig.\ref{fig:xtpmh}.
Fig.\ref{fig:sbotstop110} and \ref{fig:sbotstop140} correspond to low and 
high $\tb$ regimes respectively.  It is evident from Fig.\ref{fig:sbotstop110} 
that for low $\tb$, the effect of $A_b'$ towards determining the 
fate of the DSB vacuum is negligible and the vacuum stability is 
primarily determined by $A_t'$. This is consistent with the  
discussion of the scalar potential of NHSSM (see Sec.\ref{NHSSM}).
Both the plots identify a central region near $A_t'=0=A_b'$.  This region is associated with 
absolute stable DSB vacuum, shown in green in the plots. Bordering the 
green central region, there exists a small zone associated with
long-lived DSB minima\footnote{In some cases, {\tt Vevacious} fails to determine
the fate of the DSB vacuum when thermal corrections
to the effective potential is switched on. Without thermal corrections, these regions appear
as long-lived in our scan.}. 
 \begin{figure}[t]
 \begin{center}
 \subfigure[]{%
 \label{fig:sbotstop210}
 \includegraphics[width=0.51\textwidth]{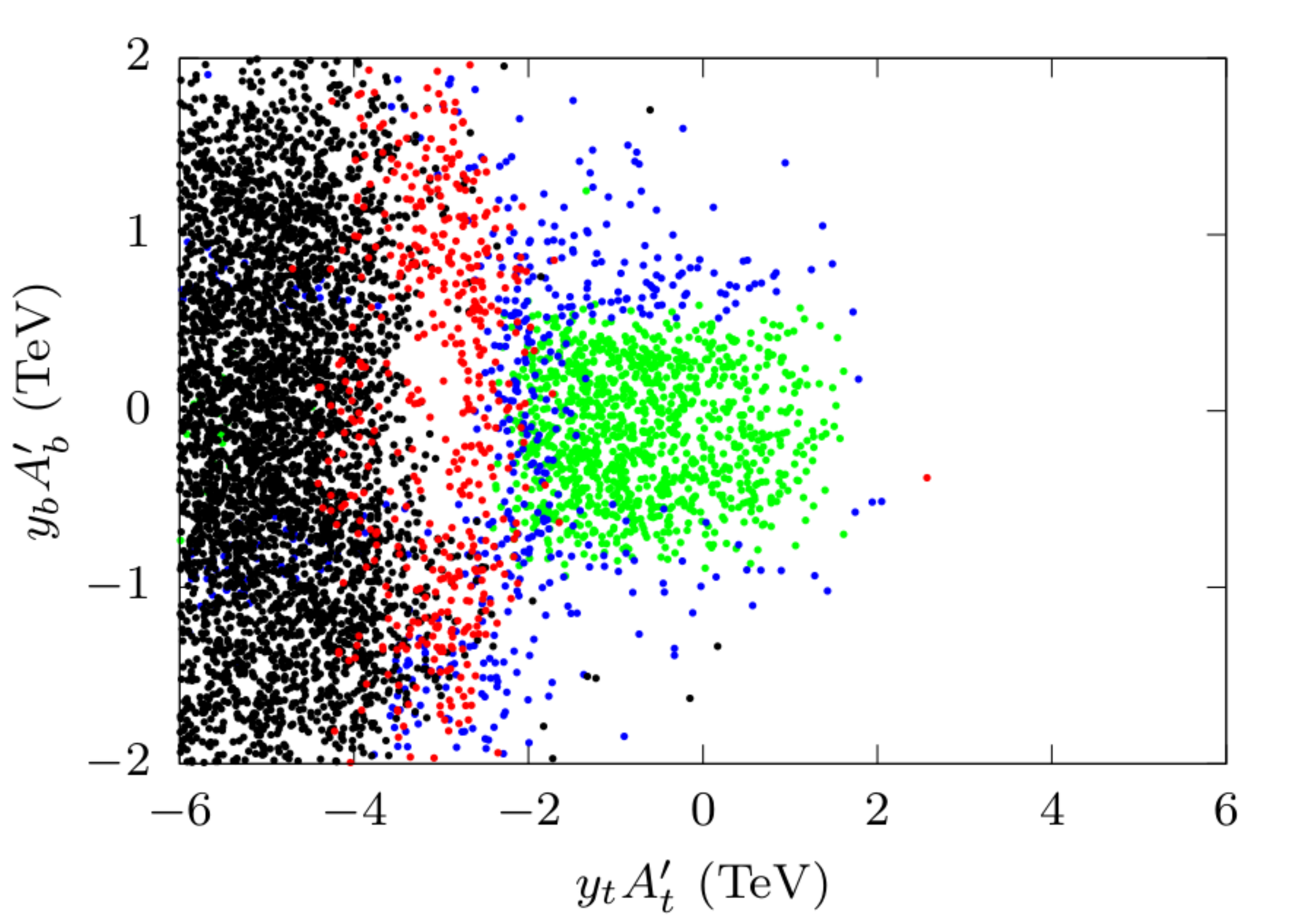}
 }%
 \subfigure[]{%
 \label{fig:sbotstop240}
 \includegraphics[width=0.51\textwidth]{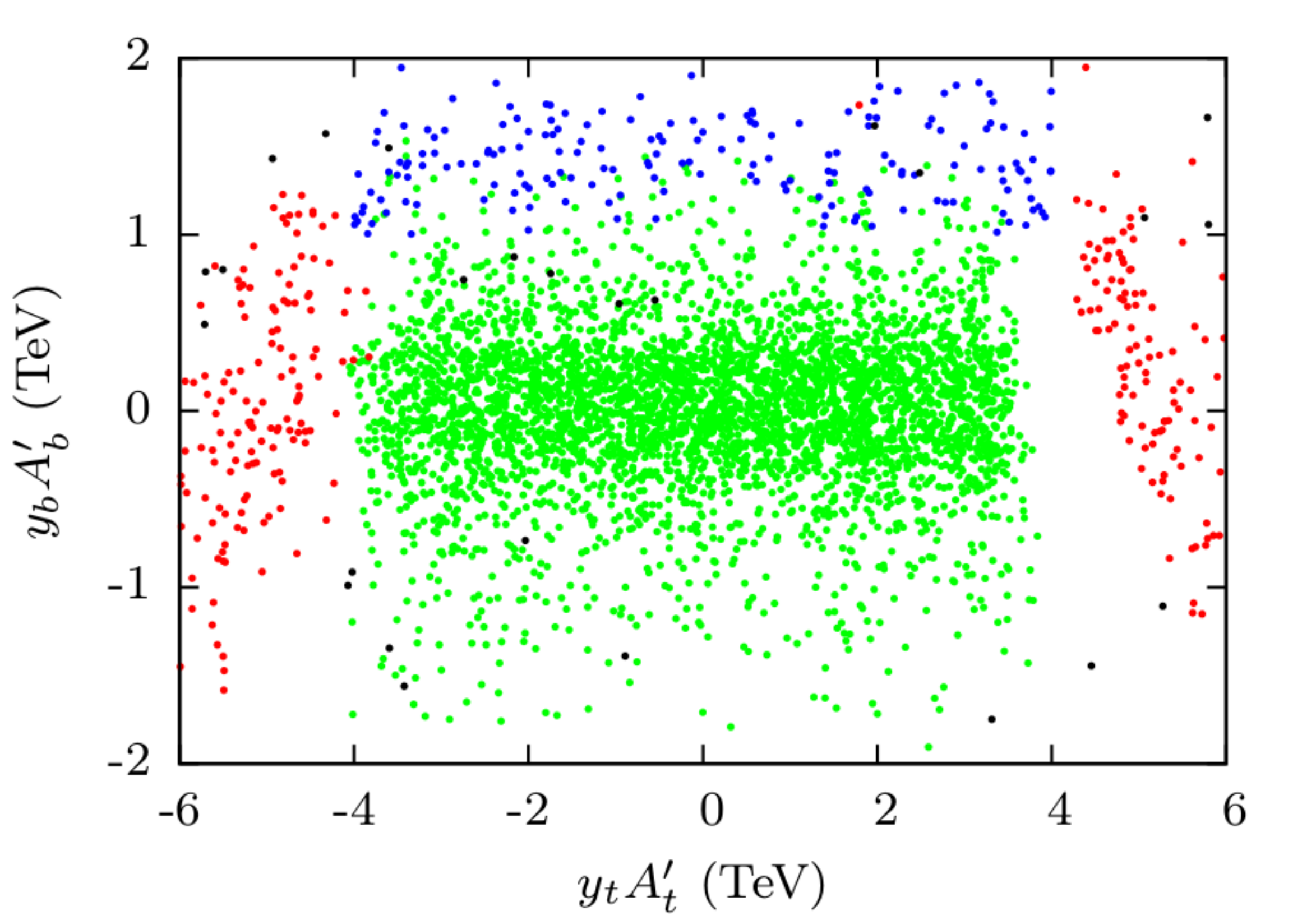}
 }%
 \caption{\it Figs.\ref{fig:sbotstop210} and \ref{fig:sbotstop240} 
 are same as Fig.\ref{fig:sbotstop1} with the limits on $m_{h_1}$ imposed. 
 It is evident that the region with $y_tA_t' > 2$ TeV is excluded via 
 limits on $m_{h_1}$ for low $\tb$.}        
 \label{fig:sbotstop2}
 \end{center}
 \end{figure}
Beyond that, there exist regions 
characterized by DSB vacua that are rendered unstable via quantum tunneling or thermal 
effects. Despite the similarity in the nature of the two 
plots, there are some striking differences that arise due to the two different 
ranges of $\tb$ that are considered in our analysis.
Fig.\ref{fig:sbotstop110} constrains $|y_tA_t'|$ to be within $2.0$ TeV for stable points, 
whereas there is almost no restriction imposed on $y_bA_b'$ when safe vacua (both stable and long-lived 
points) are considered. 
Fig.\ref{fig:sbotstop140} reveals 
that the stable points can appear for wider range of $|y_t A_t'|$ compared to that in the low $\tb$ 
limit, since its effect is more prominent for low $\tb$. 
Thus, the zone $|y_t A_t'| < 4$ TeV is easily accommodated in the model.
Demanding absolute stability one requires $|y_b A_b'| \lsim 1.5$ TeV, however, that may be relaxed by 
a small amount  via inclusion of long-lived states. We further notice that 
the metastability of the DSB vacuum is mostly governed by large $y_t A_t'$.
However, there are metastable points (mostly long-lived) even for smaller
$y_t A_t'$. The origin of such metastable points is attributed to sbottom \Vevs.
 
Now we impose the experimental limits for $m_{h_1}$ while considering a 
 $3$ GeV window resulting into the following range \cite{Degrassi:2002fi,Allanach:2004rh,
 	Martin:2007pg,Harlander:2008ju,Heinemeyer:2011aa,Arbey:2012dq,Chakraborti:2012up}.
 \begin{eqnarray}
 \label{higgsrange}
 122.1 ~\mathrm{GeV}~\leq~m_{h_1}~\leq~128.1 ~\mathrm{GeV}. 
 \end{eqnarray}
The above uncertainty essentially arises from scale dependence,
problems in computing higher order loop and renormalization
scheme related dependencies.  Imposing the above constraint of 
Eq.\ref{higgsrange} on Fig.\ref{fig:sbotstop1}, results into 
Fig.\ref{fig:sbotstop2}. Fig.\ref{fig:sbotstop210} shows that a large region of parameter 
space associated with $y_tA_t'>2$ TeV is excluded via the limits 
on $m_{h_1}$. As expected, $A_t'$ plays
a significant role in the radiative corrections to $m_{h_1}$ for 
low $\tb$. For large $\tb$, as shown in Fig.\ref{fig:sbotstop240}, 
$A_b'$ is also important in regard to vacuum stability but the radiative corrections to $m_{h_1}$ keeps it well within the range of Eq.\ref{higgsrange}. This is consistent with the results obtained from fig.\ref{fig:sbot40}. 
\section{Conclusion}
\label{conclusion}
Even after the first few years of running of the LHC, SUSY signal is yet to be observed. This has severely constrained many models of low
scale SUSY including the MSSM.  In the post Higgs@125 GeV era, the requirement of the lighter SM like Higgs boson of the MSSM to have a mass of $m_{h_1} \sim 125$ GeV translates into the need of large radiative corrections. This on the other hand,
necessitates heavier stop squark sector or large stop mixing trilinear soft parameter ($A_t$). At the same time, LHCb results on B-physics 
processes like $\bsg$ further constrain the allowed parameter space of the model especially for large $\tb$.  Regarding the DM perspective, a bino like DM requires a strong correlation among unrelated SUSY parameters to provide with the  
proper relic abundance. A wino like DM, consistent with PLANCK data, on the other hand, demands the LSP to be extremely heavy ($> 2$ TeV). 
A higgsino like DM does not require correlation between unrelated SUSY parameters and is able to satisfy the relic density limits for a relatively lower mass ($\mu \sim 1$ TeV). However, a large $\mu$ leads to a significant enhancement of EWFT. The Non-Holomorphic MSSM (NHSSM) is one of the simplest extensions of the MSSM 
that can significantly or at least partially ameliorate all of the above mentioned problems. The NHSSM
is constructed via inclusion of non-holomorphic (NH) SUSY breaking 
soft trilinear interactions with the coefficients $A_t', A_b'$ etc. along with a NH higgsino mass soft parameter ($\mu'$). In this model, $A_t'$ and to some extent $A_b'$ 
contribute to the radiative corrections to $m_{h_1}$ arising out of the stop and sbottom loops. Hence,  comparatively smaller values of stop squarks or $|A_t|$
can satisfy the higgs mass data. B-physics constraints are also more
 easily satisfied through left-right mixing in the stop sector. 
The higgsino mass in the NHSSM is governed by $|\mu-\mu'|$, whereas EWFT has no $\mu'$ dependence at the tree-level.  As a result, a higgsino like DM that provides 
required amount of relic abundance is obtained even for a lower value of $\mu$. Thus, electroweak scale naturalness is comfortably restored in the NHSSM. Although the higgs scalar potential is identical to that of the MSSM, the full
scalar potential that includes the squarks and sleptons, is modified by
the presence of NH trilinear parameters. `$\mu'$' on the other hand has no effect on the scalar potential at the tree level. The presence of charged
and colored scalars may give rise to global charge and color breaking (CCB) minima. This is a well known problem and studied widely in literature for models like the MSSM and the NMSSM. Analytic constraints
are often used to exclude the presence of global CCB minima in the MSSM. These constraints consider absolute stability of the Desired Symmetry Breaking
(DSB) minima and confine the allowed ranges of $A_t$ and $A_b$.  Consideration of
a quantum mechanically as well as thermally stable DSB vacua in presence of
a global CCB minima extends the allowed parameter space significantly.
These long-lived states along with the absolute stable ones are collectively referred to as safe vacua. 
Since the analytic constraints
are usually derived under highly simplified assumptions and are based on the choice of directions within the multi-dimensional field space, numerical analysis is required for precise understanding of the vacuum stability over the parameter space.  In this work, we first study the analytic and numerical ways of imposing the CCB constraints and study the vacuum stability by including the possibility of having long-lived states.

In the first part, we derive the analytic CCB constraints in the NHSSM. Non-vanishing \Vevs~ are attributed to the Higgs and stop fields while deriving the CCB constraints involving stop fields. On the other hand, Higgs and sbottom fields were considered to have non-zero
\Vevs~ while deriving the CCB relations in a scenario where the charge and color breaking is caused by the sbottom fields.  In a simplified scenario, all the non-vanishing \Vevs~ are considered to be equal while deriving the analytic expressions that
constrain $A_t'$ and $A_b'$ for assuring the absolute stability of the DSB minima.
However, these analytic constraints have limitations since the assumption
of equality of \Vevs~is hardly a reality in a SUSY model. 
Thus, the dependence on $\tb$ becomes quite important.  Furthermore, the analytic constraints would not be sufficient once one includes cosmologically long-lived state. 
      
Along with the theoretical analysis, we follow a semi-analytic approach using complete 1-loop corrected effective potential implemented via {\tt Mathematica}
to investigate the influence of the NH terms on the scalar potential of the model. We find that, the terms associated with $A_t'$
and $A_b'$ effectively lower the scalar potential of the model. This
lowering essentially leads to the appearance of global CCB minima depending on the values of the NH parameters. We also present an  exhaustive scan over relevant
region of parameter space by using the dedicated package {\tt Vevacious} that includes both zero temperature effects and finite temperature corrections to the NHSSM potential. 

We divide our {\tt Vevacious} based analysis into three different parts
depending on the fields whose \Vevs~ are allowed to be non-vanishing.
First, in order to probe the effect of the stop fields, non-vanishing \Vevs~
 are attributed to stops and higgs fields.  The effect of $A_t'$ is more
 prominent for low $\tb$, where the former is confined in a quite small
range via the requirement of safe vacua. On the other hand, for large
$\tb$, a comparatively wider range of $A_t'$ is allowed. This is due to the fact, that the contribution of $A_t'$ in the scalar potential is $\tb$ suppressed. The region of
parameter space with adequate value of $m_{h_1}$, is associated with both safe and dangerous vacua. $m_{h_1}$ becomes important in a CCB related study, simply because a right zone of mass of the higgs boson require large 
values of trilinear parameters $A_t$ and $A_t'$ that are sensitive to CCB analyses. The
results obtained from {\tt Vevacious} based analysis show the presence of
a significant region of long-lived states. Hence the region of parameter space 
having safe vacua can be extended considerably which is otherwise ruled out by
analytical CCB constraint. 

Exploring the model for CCB minima associated with the sbottom fields,
we find that the CCB constraint is hardly effective in constraining 
$A_b'$ for small values of $\tb$. For large $\tb$, there are long-lived
states even if the analytical CCB constraint is satisfied. We also find that
most of the region is safe against deeper CCB vacua arising from sbottom \Vevs.

In a more realistic scenario, stop and sbottom fields together modify the
scalar potential. In order to probe this effect, all the third 
generation squarks along with the higgs fields are assumed to have
non-vanishing \Vevs. We try to locate the safe vacuum in the
$y_t A_t'-y_b A_b'$ plane. The origin ($A_t'= A_b'= 0$) 
corresponds to the MSSM scenario where the NH terms are absent. 
There exists a region of absolute stable vacuum around $A_t'=0=A_b'$. This zone is 
bounded by regions that are characterized by the presence of long-lived DSB minima. 
Beyond this, almost the entire $y_t A_t'-y_bA_b'$ plane 
is excluded due to the instability of DSB minima. 
A wider range of $y_tA_t'$ is allowed for large $\tb$, but it
is confined in a quite smaller range for low $\tb$. Moreover,
the allowed range of $m_{h_1}$ has visible
impact for the case with low $\tb$ and $A_t'>0$. On the other hand,
most of the region with large $\tb$ is consistent with such a range.
\section{Acknowledgement}
AD would like to thank Indian Association for the 
Cultivation of Science for infrastructural support.  
JB  is  partially  supported  by  funding  available  from  
the  Department  of  Atomic  Energy, Government of India for 
the Regional Centre for Accelerator-based Particle Physics (RECAPP), 
Harish-Chandra Research Institute. JB is also partially supported by the `INFOSYS
scholarship for senior students'. The authors acknowledge AseshKrishna Datta
and Utpal Chattopadhyay for their useful comments and encouragement. 
The authors also acknowledge the use of the cluster computing setup 
available at the High-Performance Computing facility of HRI.
			
\end{document}